\documentclass[letterpaper]{article} % DO NOT CHANGE THIS
\usepackage{aaai25}  % DO NOT CHANGE THIS
\usepackage{times}  % DO NOT CHANGE THIS
\usepackage{helvet}  % DO NOT CHANGE THIS
\usepackage{courier}  % DO NOT CHANGE THIS
\usepackage[hyphens]{url}  % DO NOT CHANGE THIS
\usepackage{graphicx} % DO NOT CHANGE THIS
\usepackage{natbib}  % DO NOT CHANGE THIS AND DO NOT ADD ANY OPTIONS TO IT 
\usepackage{caption} % DO NOT CHANGE THIS AND DO NOT ADD ANY OPTIONS TO IT
\usepackage{cleveref}
\usepackage{booktabs}
\usepackage{array}
\usepackage[utf8]{inputenc}
\usepackage{enumitem}
\usepackage{amssymb}
\usepackage{comment}

\newcounter{questionCounter}
\newenvironment{survey}{}{}
\newenvironment{question}[1][]
    {\par\stepcounter{questionCounter}
    \noindent\textbf{q\thequestionCounter. #1}}
    {\par}
\newenvironment{choices}
    {\begin{enumerate}[label=\arabic*., leftmargin=*]}
    {\end{enumerate}}
\newenvironment{checkboxes}
    {\begin{enumerate}[label=$\square$, leftmargin=*, labelsep=0.5em]}
    {\end{enumerate}}
\newenvironment{scale}[1]
    {\par\noindent\textit{Scale:}\begin{enumerate}[label=\arabic*., leftmargin=*]#1\end{enumerate}
    \par\noindent \textit{Subquestions:}\begin{enumerate}[label=\alph*., leftmargin=*]}
    {\end{enumerate}}

\usepackage{algorithm}
\usepackage{algorithmic}

\usepackage{newfloat}
\usepackage{listings}
\DeclareCaptionStyle{ruled}{labelfont=normalfont,labelsep=colon,strut=off} % DO NOT CHANGE THIS
\floatstyle{ruled}
\newfloat{listing}{tb}{lst}{}
\floatname{listing}{Listing}
\title{Responsible AI in the Global Context: Maturity Model and Survey}
\author {
    Anka Reuel\textsuperscript{\rm 1, *},
    Patrick Connolly\textsuperscript{\rm 2, *},
    Kiana Jafari Meimandi\textsuperscript{\rm 1}
    Shekhar Tewari\textsuperscript{\rm 2},
    Jakub Wiatrak\textsuperscript{\rm 2},
    Dikshita Venkatesh\textsuperscript{\rm 2},
    Mykel Kochenderfer\textsuperscript{\rm 1}
}
\affiliations {
    \textsuperscript{\rm 1}Stanford University\\
    \textsuperscript{\rm 2}Accenture\\
    \textsuperscript{\rm *}Co-first authors\\
    Corresponding author: anka.reuel@stanford.edu
}

\usepackage{bibentry}
\begin{document}
\maketitle

\begin{abstract} 
Responsible AI (RAI) has emerged as a major focus across industry, policymaking, and academia, aiming to mitigate the risks and maximize the benefits of AI, both on an organizational and societal level. This study explores the global state of RAI through one of the most extensive surveys to date on the topic, surveying 1000 organizations across 20 industries and 19 geographical regions. We define a conceptual RAI maturity model for organizations to map how well they implement organizational and operational RAI measures. Based on this model, the survey assesses the adoption of system-level measures to mitigate identified risks related to, for example, discrimination, reliability, or privacy, and also covers key organizational processes pertaining to governance, risk management, and monitoring and control. The study highlights the expanding AI risk landscape, emphasizing the need for comprehensive risk mitigation strategies. The findings also reveal significant strides towards RAI maturity, but we also identify gaps in RAI implementation that could lead to increased (public) risks from AI systems. This research offers a structured approach to assess and improve RAI practices globally and underscores the critical need for bridging the gap between RAI planning and execution to ensure AI advancement aligns with human welfare and societal benefits. 
\end{abstract}

\section{Introduction}\label{sec:intro}
Responsible AI (RAI) has become a central theme in discussions about AI to mitigate risks and harness the benefits of the technology. Recent surveys highlight the growing importance of RAI for organizations worldwide. For instance, a recent report reveals that 52\% of companies engage in some level of RAI, albeit with limited scale and scope in most cases~\cite{mitsloan2022}.
The need for understanding RAI adoption on a company level is increasingly important: RAI measures help identify potential weaknesses that can lead to AI-related harms such as bias, discrimination, and privacy breaches. These issues carry significant societal consequences, potentially exacerbating existing inequalities if not mitigated through widespread organizational and system-level adoption of RAI practices. The absence of standardized metrics impedes the assessment of RAI progress and, consequently, the effective management of AI's societal impact. Regulatory pressure, exemplified by the EU AI Act mandating pre-deployment evaluations for certain AI systems, reflects growing societal concerns about the technology's potential negative effects and necessity for organizations to demonstrate RAI adoption. Demonstrating adherence to RAI practices further builds stakeholder trust and contributes to AI's social acceptance. However, despite its importance, a lack of consensus persists on a standardized, scalable approach to evaluate RAI within organizations.
Most existing work ~\cite{xia2023towards,wang2023designing} focuses on system-level aspects and overlooks organizational considerations, presenting an incomplete picture of RAI adoption, given that organizational processes and plans are a key aspect to ensuring consistency in RAI implementation across different AI projects and teams. This gap in the literature limits our understanding of how RAI principles and frameworks are operationalized and hinders the development of comprehensive strategies for improving RAI practices. Furthermore, the lack of attention to organizational factors may lead to a disconnect between high-level RAI principles and their practical application in day-to-day operations, potentially undermining the effectiveness of RAI initiatives.

This work addresses these challenges by developing a conceptual RAI maturity model for organizational and system-level RAI maturity. Based on this model, we measure current levels of RAI maturity in a first-of-its-kind global survey with 1000 organizations across 20 industries and 19 geographical regions to track progress and identify areas for improvement in RAI adoption. As part of the survey, we further capture aspects like technical risk mitigation measures, risk perceptions, implemented AI governance structures, and barriers to generative AI adoption to gauge current perceptions regarding the responsible development, deployment, and use of AI, shedding light on how these perceptions might influence RAI maturity. Our aim extends beyond mere quantification of adoption rates. We sought to explore the correlations between RAI adoption, risk exposure, regulatory exposure, and organizations' positions within the supply chain in relation to their RAI activities.

The rest of this work is structured as follows: We outline the background to our work in Sec.~\ref{sec:back}, our RAI maturity model in Sec.~\ref{sec:raimm}, and our survey methodology in Sec.~\ref{sec:method}. We continue to present the survey results in Sec.~\ref{sec:result}, before discussing their context and implications in Sec.~\ref{sec:dis} and limitations in Sec.~\ref{sec:limit}.

\begin{table*}[h!]
    \centering
    \small
    \begin{tabular}{>{\raggedright\arraybackslash}m{3.5cm} >{\raggedright\arraybackslash}m{11cm} > {\raggedright\arraybackslash}m{2cm}} 
        \toprule
        \textbf{Component} & \textbf{Description} & \textbf{Question}\\
        \midrule        \textbf{Reliability} & Necessitates that performance is reliable across all relevant (sub-) groups and that AI systems are resilient to attacks, and equipped with safety measures and fallback plans. & Q32 \\
        \textbf{Privacy \& Data Governance} & Necessitates respect for privacy, data quality, and cybersecurity, including appropriate access controls. & Q29 \& Q35\\
        \textbf{Human Interaction} & Protection of fundamental human rights, limitation of misuse, and measures to prevent overreliance on AI systems & Q28\\
        \textbf{Transparency} & AI system that is explainable to the technical user and interpretable to the end-user; also includes release of details about the training data, architecture design, and other relevant system information. & Q33\\
        \textbf{Societal \& Environmental Wellbeing} & Promotion of sustainability, positive social contributions, and democratic values. & Q31 \\
        \textbf{Diversity, Non-discrimination \& Fairness} & AI systems must be developed in a way to actively avoid unfair bias, ensuring accessibility and inclusivity for diverse users through universal design principles and active stakeholder participation in the design process. & Q30 \\
        \textbf{Accountability} & Mandating auditability of systems, proactive measures to minimize negative impacts, transparent trade-off discussions, along with mechanisms for recourse and remediation in the event of any adverse effects. & Q34 \\
        \bottomrule
    \end{tabular}
    \caption{RAI system-level/operational dimensions, based on \citet{european_commission2019high}. Specific RAI measures for each component are listed in App.~\ref{app:survey_questions} under the  question referenced here.}
    \label{tab:raiop_dimensions}
\end{table*}

\section{Background}\label{sec:back}
\textbf{Responsible AI.} Even though RAI is considered a critical aspect in the current AI ecosystem, the concept of RAI lacks a universally accepted definition~\cite{wang2023designing}. Many definitions emphasize the practices used to implement RAI rather than providing a conceptual understanding of its essence~\cite{wang2023designing,xia2023towards}. In this work, we base our RAI definition on that of the European Commission’s High-Level Expert Group definition of AI principles for system-level RAI dimensions~\cite{highlevelexpertgroup2018}. Their work outlines key aspects of RAI across well-defined categories~\cite{pekka2018european}, was developed by a group of subject-matter experts, and was pre-step to the newly enacted EU AI Act. It is also recognized by various organizations~\cite{add}.

The European Commission's understanding of RAI encompasses a set of seven system-level (or operational) principles aimed at ensuring ethical and trustworthy AI. We adapted them to account for new developments in the field since the publication of the European Commission’s work in 2018 and used the set of system-level RAI dimensions listed in Tab.~\ref{tab:raiop_dimensions} in our research. The European Commission's definition does not include organizational-level aspects in their RAI understanding. In line with the human-centered AI governance model of \citet{shneiderman2020bridging}, we emphasize the importance of organizational processes and governance frameworks, as well as aspects such as training, culture, and leadership support to drive RAI within a company. We add the following dimensions to the RAI understanding used in this study:

\begin{itemize}
    \item \textit{Organizational Governance}, encompassing structures, processes, and roles within an organization to support RAI efforts~\cite{pwc2019practical,pwc2021maturing}.
    \item \textit{Leadership and Culture}, including support from C-level executives, responsibility-first culture, and RAI training efforts~\cite{pwc2019practical,pwc2021maturing}.
\end{itemize}

This proposed set of combined dimensions serves as a foundation for the development of our RAI maturity model and RAI survey. It is important to note that these dimensions can often overlap and interact. For instance, transparency is crucial for ensuring fairness and system integrity. Similarly, accountability mechanisms support both reliability and transparency, and in part rely on organizational governance measures to be addressed.

\textbf{RAI Questionnaires.} In recent years, there has been a surge in corporate and governmental efforts to assess RAI. Existing RAI questionnaires and frameworks vary widely, developed by corporations~\cite{pwc2023diagnostics,initiative_for_applied_ai2023maturity}, governments~\cite{ai2023artificial,pekka2018european}, and other organizations~\cite{responsible_ai_institute2021certification,ai_global2020responsible}. These questionnaires differ in type (e.g., context-specific assessment, survey, certification), number of questions, high-level RAI dimensions covered, and methodology. Despite this growing interest, significant gaps exist in current RAI questionnaires and models. Many lack transparent methodologies, fail to define RAI terms or the dimensions they cover, and offer unclear scoring and interpretation. Furthermore, these questionnaires often fail to assess both system-level, i.e., operational, \textit{and} organizational aspects of RAI.

To address the gaps, we specify our definition of the components of RAI, along with the survey structure and maturity logic, and scoring. We further ensure that both system-level mitigation measures and organizational RAI practices are captured by our survey to present the most comprehensive, public RAI survey to date. We further recognize the need for global comparability and have adopted a questionnaire-based approach, which enabled us to scale the survey and gather data from over 1000 organizations worldwide. 

\section{RAI Maturity Model}\label{sec:raimm}
As part of the development of our RAI survey, we recognized the lack of and need for an in-depth RAI maturity model that captures how well organizations adopt organizational and system-level, i.e., operational RAI measures and mitigate risks associated with the development, deployment, and use of AI. Maturity models are a well-established tool across fields \cite{wendler2012maturity}, with models being developed for software management \cite{john2021towards, paulk1994comparison}, project management \cite{crawford2021project}, and business processes \cite{lee2007overview, hillson1997towards, fisher2004business}. For AI, there have been a number of maturity models proposed \cite{sadiq2021artificial, fukas2021developing, alsheibani2019towards, ellefsen2019striving}. However, while there has been interest in a maturity model with a focus on responsible AI \cite{vakkuri2021time}, existing efforts have been incomplete. Some are limited to system-level RAI maturity models, without accounting for organizational RAI measures \cite{jantunen2021building}. Other efforts have only reviewed literature on (R)AI maturity models without developing their own model based on the identified gaps \cite{akbarighatar2022maturity, akbarighatar2023sociotechnical}. Previous studies \cite{akbarighatar2022maturity, lichtenthaler2020five, mukherjee2022analysing} emphasize the importance of both technical and organizational aspects in RAI maturity, but they do not develop a corresponding comprehensive maturity model themselves. A detailed maturity model is necessary to guide organizations in adopting AI responsibly. This work aims to fill this gap. In addition, we are the first to apply our model in practice, surveying 1000 organizations globally based on our maturity model.

The model comprises two main dimensions: organizational and operational RAI maturity, each with distinct subcomponents and corresponding maturity levels. We iteratively refined both the main and subcomponents and the maturity levels for both based on a synthesis of the literature we outlined above and expert interviews with stakeholders in academia, governance, and industry. The level namings are consistent with standard maturity model naming conventions, as used for example in \citep{fukas2021developing} or \citep{alsheibani2019towards}. The model presented in Tab.~\ref{table:maturitymodel}, along with the maturity levels defined for the subcomponents (see App.~\ref{app:subcomponent_maturitylvls}), can serve as a tool for organizations to identify areas for improvement in adopting RAI practices. It further served as the foundation for our survey design to define the components covered in the survey, as well as the answer options for each question.

\begin{table*}[h!]
    \centering
    \begin{tabular}{>{\raggedright\arraybackslash}m{1.8cm} >{\raggedright\arraybackslash}m{1.75cm} >{\raggedright\arraybackslash}m{5.25cm} >{\raggedright\arraybackslash}m{1.75cm} >{\raggedright\arraybackslash}m{5.25cm}}
        \toprule
        \textbf{Level} & \textbf{Score} & \textbf{Organizational Maturity} & \textbf{Score} & \textbf{Operational Maturity} \\
        \midrule
        \textbf{Level 1: Initial} & [0 , 12.5] & The organization has limited awareness and no organizational plans, processes, or frameworks in place to ensure a responsible AI adoption. & [0 , 12.5] & The organization does not mitigate identified risks on a system level. \\
        \textbf{Level 2: Assessing} & [12.5 , 37.5] & The organization is aware of the necessity for organizational measures to ensure a responsible AI adoption and is assessing governance options. & [12.5 , 37.5] & Awareness of risks may be present, but the organization has only limited or no formal mitigation measures in place. \\
        \textbf{Level 3: Determined} & [37.5 , 62.5] & The organization demonstrates foundational governance capabilities to support the responsible development, deployment, and use of AI. & [37.5 , 62.5] & A few risk mitigation measures are being fully operationalized, but the majority is only implemented ad-hoc or in early roll-out stages. There is a growing awareness of the need for more systematic approaches. \\
        \textbf{Level 4: Managed} & [62.5 , 87.5] & The organization has established comprehensive organizational RAI measures and is actively ensuring enterprise-wide adoption, demonstrating a mature and effective approach to internal RAI governance. & [62.5 , 87.5] & A wide range of risk mitigation measures are fully operationalized across all relevant AI systems in the organization. \\
        \textbf{Level 5: Optimized} & [87.5 , 100] & The organization demonstrates an established, future-oriented approach towards organizational RAI, ensuring a sustainable and responsible approach to organizational RAI. & [87.5 , 100] & Comprehensive, state-of-the-art risk mitigation strategies are fully operationalized. The organization continuously monitors and evaluates risks, proactively adapting its practices as needed to mitigate new risks. \\
        \bottomrule
    \end{tabular}
    \caption{Overall organizational and operational maturity model and corresponding score brackets used for our global RAI survey. Individual maturity models per subcomponent can be found in App.~\ref{app:subcomponent_maturitylvls}.}
    \label{table:maturitymodel}
\end{table*}

\subsection{Organizational RAI Maturity}\label{sec:orgmat}
Organizational RAI maturity measures the sophistication and effectiveness of  organizational processes, frameworks, and culture in ensuring its responsible development, deployment, and use of AI. Based on previous work (see Tab.~\ref{tab:raiorg_subcomponents} for references), we subsume eight subcomponents in organizational RAI maturity, summarized in Tab.~\ref{tab:raiorg_subcomponents}. Survey questions Q14, Q17, Q20–Q24, and Q33 capture these components in our survey (see App.~\ref{app:survey_questions}). Components and corresponding maturity levels were included based on a literature review for each dimension as well as expert interviews. Scoring details are discussed in Sec.~\ref{sec:dataanalysis_scoring}. 

\subsection{Operational RAI Maturity}\label{sec:opmat}
Operational RAI maturity assesses the comprehensiveness of an organization's implementation of system-level mitigation measures for identified AI adoption risks. Operational RAI maturity encompasses all system-level aspects of RAI discussed in Tab.~\ref{tab:raiop_dimensions}. In our survey, we first presented respondents with a list of potential risks (detailed in App.~\ref{app:survey_questions}), including privacy, diversity, reliability, transparency, security, human interaction, environmental impact, and accountability concerns. Respondents selected relevant risks and indicated their level of implementation for pre-defined mitigation measures for these identified risks, such as red teaming, mitigations for adversarial attacks, bias mitigation, or uncertainty quantification (questions Q22b, Q25–Q31, see App.~\ref{app:survey_questions} for all mitigation measures). Options for `Other' with a free-form text field to add not-listed mitigations measures, and `None' were also provided. Scoring details are discussed in Sec.~\ref{sec:dataanalysis_scoring}. Mitigation measures were included based on a literature review for each dimension and expert interviews. Scoring details are discussed in Sec.~\ref{sec:dataanalysis_scoring}.

\begin{table*}[t]
    \small
    \centering
    \begin{tabular}{>{\raggedright\arraybackslash}p{2.5cm} >{\raggedright\arraybackslash}p{10cm} >{\raggedright\arraybackslash}p{4cm}}
        \toprule
        \textbf{Sub Component} & \textbf{Description} & \textbf{Source}\\
        \midrule
        \multicolumn{3}{c}{\textbf{Organizational Governance}}\\
        \midrule
        Governance Operating Model & Structures, policies, and processes that guide AI system development and deployment. & \citet{lu2024responsible, batool2023responsible} \\ 
        Risk Identification & Systematic processes to recognize the potential risks and developing proactive measures to detect problems in AI systems before they escalate. & \citet{lu2024responsible, clarke2019principles, qureshi2024ai} \\
        Risk Assessment \& Management Framework & Comprehensive risk management framework and processes for the evaluation of the likelihood and potential impact of the identified risks to prioritize their management. & \citet{clarke2019principles, wirtz2022governance} \\
        Risk Mitigation & Processes to implement strategies and controls to reduce the potential negative impacts of AI systems; Includes frameworks to address potential negative impacts on individuals, the organization, and society. & \citet{clarke2019principles, qureshi2024ai} \\
        Procurement & Processes involved in acquiring AI technologies and services from external vendors; Ensuring procurement decisions align with RAI principles. & \citet{obinna2024developing} \\
        Monitoring \& Control & Ongoing oversight of AI systems to ensure they operate as intended and adhere to the guidelines and mechanisms for detecting and correcting deviations. & \citet{camilleri2024artificial} \\
        Cybersecurity & Protection of AI systems and data from cyberthreats. Implementing robust security measures to safeguard against unauthorized access, data breaches, and other cyberrisks. & \citet{lannquist12020intersection, bowen2024technological}  \\
        \midrule
        \multicolumn{3}{c}{\textbf{Leadership and Culture}}\\
        \midrule
        Sponsorship & Support and commitment from senior leadership for RAI initiatives and allocating resources, setting priorities, and endorsing RAI practices. & \citet{lu2024responsible} \\
        Training & Education and training of employees and stakeholders on RAI practices. & \cite{ }\\
        \bottomrule
    \end{tabular}
    \caption{RAI organizational dimensions and their sub components. Detailed maturity levels for each subcompontent can be found in App.~\ref{app:subcomponent_maturitylvls}.}
    \label{tab:raiorg_subcomponents}
\end{table*}

% \begin{table}[t]
%     \small
%     \centering
%     \begin{tabular}{>{\raggedright\arraybackslash}p{2cm} >{\raggedright\arraybackslash}p{4cm} >{\raggedright\arraybackslash}p{2cm} >{\raggedright\arraybackslash}p{2cm}}
%         \toprule
%         \textbf{RAI Maturity Type} & \textbf{Components} & \textbf{No. Qs} & \textbf{Sub Qs} \\
%         \midrule
%         \textbf{Operational Maturity} & Human agency & Q25 & 6 \\
%         & Privacy \& Data Governance & Q26 & 6 \\
%         & & Q32 & 5 \\
%         & Diversity, non-discrimination \& Fairness & Q27 & 5 \\
%         & Societal \& Environmental Wellbeing & Q28 & 5 \\
%         & Technical Robustness \& Safety & Q29 & 6 \\
%         & Transparency & Q30 & 4 \\
%         & Accountability & Q31 & 7 \\
%         \midrule
%         \textbf{Organizational Maturity} & Sponsorship & Q14 & 1 \\
%         & Governance & Q17 & 1 \\
%         & Risk Management & Q20 & 1 \\
%         & Risk Identification & Q21 & 1 \\
%         & Risk Mitigation & Q22a & 1 \\
%         & Monitoring \& control & Q23a & 1 \\
%         & Cybersecurity & Q24 & 1 \\
%         & Training & Q33 & 1 \\
%         & Procurement & Q22b & 7 \\
%         \bottomrule
%     \end{tabular}
%     \caption{Components and their associated questions and sub-questions used in the RAI maturity assessment.}
% \end{table}

\section{Methodology}\label{sec:method}
\subsection{Questionnaire Design}\label{sec:qd}
The survey included 39 questions covering system- and organizational-level dimensions of RAI. Ten were qualifier questions about job function, company location, global revenue, visibility into RAI decision-making, industry, and general AI adoption strategy. Respondents had to indicate whether their organization was developing AI models, modifying third-party models, or using third-party models without modification. They also specified if these models were intended for internal use only or for resale. The remaining 29 questions addressed governance, risk management, system-level RAI mitigation measures, talent availability, lawfulness and compliance, and generative AI. The questionnaire was adapted based on responses to the qualifier questions (see App.~\ref{app:survey_questions}). We also considered interviews as a mode for data collection but decided to use a survey to collect data from a broader sample for global representation.

\subsection{Data Collection}\label{sec:qc}
To ensure the relevance of the responses, we included only C-suite respondents who were directly involved in or had visibility into RAI-related strategies and priorities. We targeted companies actively developing, selling, or using AI commercially, focusing on those identifying specific AI risks or anticipating AI regulation within five years. Companies with less than \$499 million USD in global annual revenue or where respondents lacked visibility into the RAI decision-making process were excluded. 
An outside firm was contracted to manage recruitment and data collection, using a double opt-in process and various recruitment channels such as LinkedIn and partnerships. Quality control measures were implemented to ensure data integrity, including screening for low-quality responses, which were identified by unusually short completion times or patterns of uniform answers.
The study was designed to ensure global representation by selecting 20 countries across all major geographical regions. We received 1,000 complete responses across 19 industries. To accommodate non-English-speaking respondents, surveys were translated into official languages of the respective countries. Data collection occurred between February and March 2024, with participation being anonymous and voluntary. Informed consent was obtained, and respondents had the option to withdraw at any point. Participants were incentivized through the external firm's incentive system.

\subsection{Data Analysis \& Scoring}\label{sec:dataanalysis_scoring}
For questions assessing organizational processes and frameworks (Q14, Q17, Q20–Q24, Q33), we formulated the answer options in our survey so that they corresponded, for the respective subcomponent assessed, to the organizational maturity levels outlined in Tab.~\ref{table:maturitymodel}, with each level scoring 0, 25, 50, 75, or 100 points respectively for each subcomponent (see App.~\ref{app:subcomponent_maturitylvls} for all subcomponent-specific maturity levels). The overall organizational RAI maturity score was derived by averaging the scores from these questions, providing a measure of a company’s organizational RAI maturity. For questions assessing concrete measures taken within a system-level RAI dimension (Q22b, Q25–Q31), we assigned 100 points if a mitigation measure has been indicated to be fully operationalized and 0 otherwise. We only asked about measures for risks that the respondents previously identified as relevant to their AI adoption. For operational RAI maturity, we averaged the points across all relevant mitigation measures and assigned a maturity level based on the resulting average score (see Tab.~\ref{table:maturitymodel} for scoring brackets). The scores were further aggregated at multiple levels (e.g., region, industry) to enable comparative analyses (see App.~\ref{app:add_results}). Independent samples t-tests ($\alpha = 0.05$) were employed to assess statistical significance of group differences. We further ran regression analysis between selected questions (see App.~\ref{app:regression_results}).

\section{Results}\label{sec:result}
\subsection{Maturity}
\subsubsection{Organizational RAI Measures}\label{sec:orgrai}
\begin{figure}
    \centering
    \includegraphics[width=\linewidth]{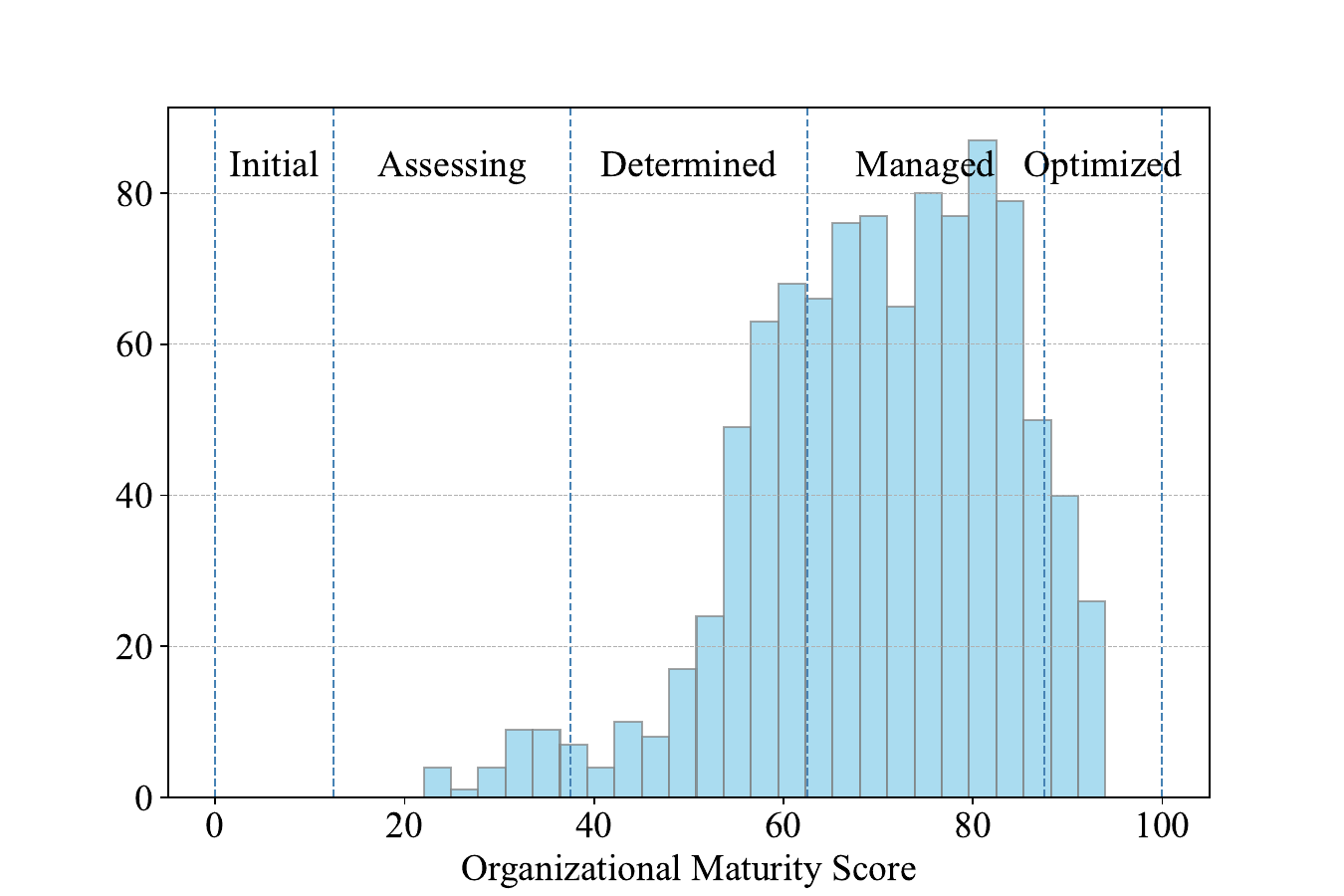}
    \caption{Organizational RAI maturity distribution.}
    \label{fig:org_dist}
\end{figure}

Most organizations are at the \textit{Managed} stage of organizational RAI maturity, with only 9\% reaching the \textit{Optimized} stage, indicating that fully integrated RAI practices are still uncommon. However, the majority of organizations are at mid-level organizational RAI maturity which can be interpreted as an indication of a broad recognition of RAI while highlighting challenges in advancing beyond this stage (Fig.~\ref{fig:org_dist}). 
Maturity varies by region, with North America and Asia leading, particularly Singapore and Japan. Latin America lags, with many organizations at the \textit{Initial} stage and the lowest average organizational RAI maturity score (see App.~\ref{app:add_results}). 
Across the selected 19 industries, \textit{Healthcare}, \textit{Life Sciences}, and the \textit{Communication, Media, \& Technology} sectors exhibit high RAI maturity, likely driven by higher regulatory demands and technological innovation within these industries. In contrast, \textit{Natural Resources} and \textit{Utilities} show lower maturity, highlighting a need for improvement (see App.~\ref{app:add_results}).

\subsubsection{Operational RAI Measures}\label{sec:oprai}
The operational RAI index results in a right-skewed distribution, with a mean score around 35. A floor effect shows 7\% of respondents scored 0, indicating no operationalized RAI measures. The majority of organizations are at a mid-level of operational maturity while only a small fraction of organizations reach the \textit{Optimized} stage, indicating that achieving full operational maturity in RAI is challenging and potentially under-prioritized. Regionally, mean operational maturity scores are similar, suggesting uniform global operational RAI maturity. However, Latin America has the highest proportion of companies with no operationalized RAI measures, indicating a lag in adoption. North America and Europe show the lowest proportions, reflecting more widespread RAI implementation and a higher baseline maturity.

\subsection{Individual Questions}
\subsubsection{Risk Perception}\label{sec:risk}
The top risks associated with AI were privacy and data governance~(51\%), cybersecurity~(47\%), and reliability~(45\%) concerns. These highlight the importance of managing sensitive data and ensuring secure, robust AI operations. In contrast, non-discrimination and fairness~(29\%) and accountability~(26\%) concerns were seen as less urgent. Organizational/business~(12\%) and brand risks~(26\%) were among the least selected, indicating a lower perceived impact on operations and direct concern for reputation. Geographically, organizations vary in the number of risks they identify, with Asian companies selecting the most~(4.99), followed by Rest of the World~(4.56), Europe~(4.32), and North America~(4.19), reflecting regional differences in risk perception. A significant majority~(88\%) of organizations believed the responsibility for mitigating risks specifically from generative AI  lies with foundation model developers, not end-users. Additionally, human interaction risks~(35\%) and environmental issues~(30\%) were increasingly recognized, reflecting growing concerns about the societal and ecological impacts of AI.

\subsubsection{View of RAI}\label{sec:raiview}
\begin{figure*}[]
    \centering
    \includegraphics[width=0.7\linewidth]{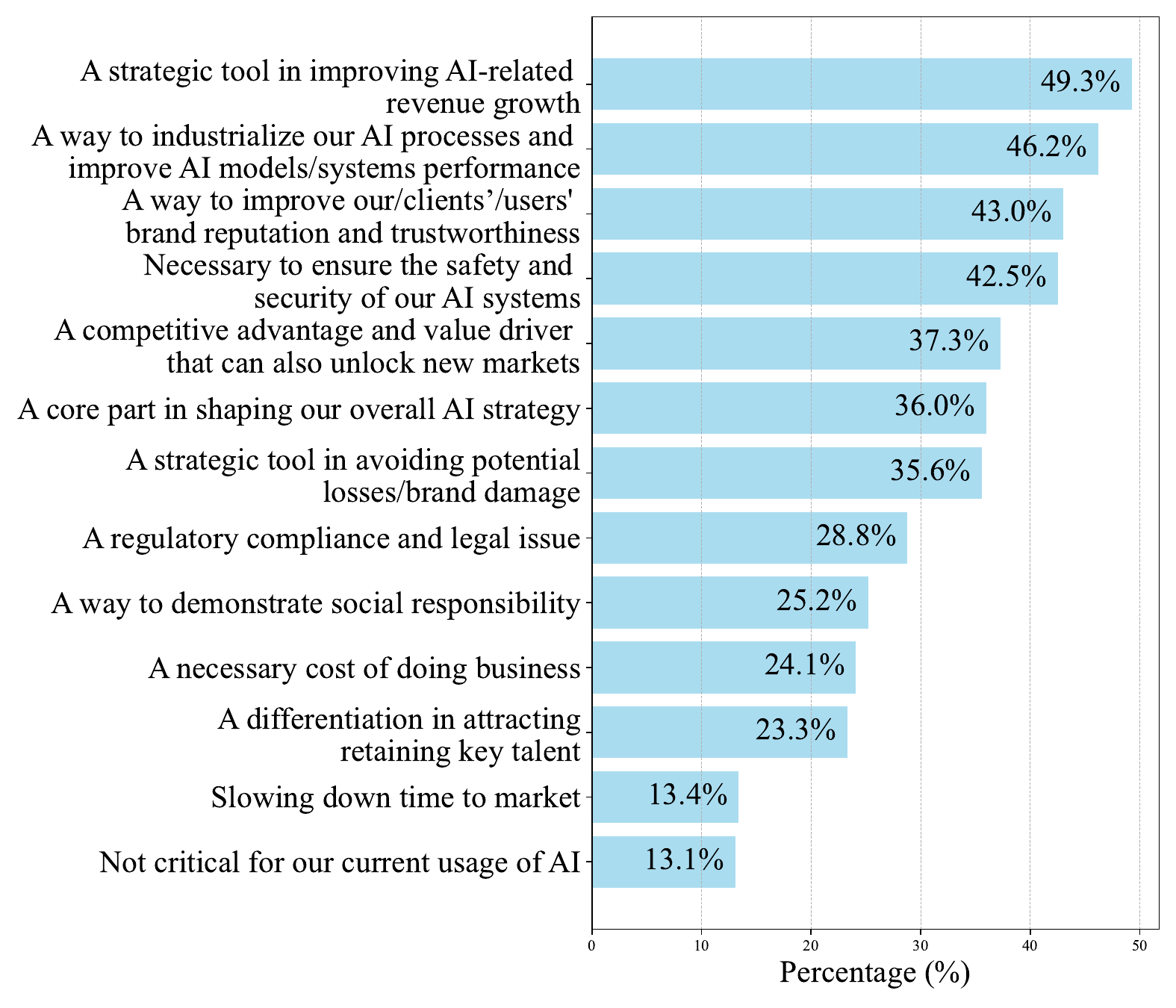}
    \caption{The way RAI is viewed within  organizations.}
    \label{fig:rai_view}
\end{figure*}

Participants were asked about how they view RAI within their organization (Fig.~\ref{fig:rai_view}). Out of all respondents, 49.3\% indicated that they see RAI as a strategic tool for revenue growth while 46.2\% answered that it is a way to improve the performance of their AI models/systems. Furthermore, 43\% saw it as a way to improve their brand reputation and trustworthiness, and 42.5\% saw it as a necessity to ensure the safety and security of their AI systems. In comparison, 13.1\% indicated that it was not critical for their current usage of AI systems, while only 13.4\% indicated it is slowing down innovation and time to market. These results highlight that overall, participants saw RAI as a value driver, and only a minority indicated that it would not be relevant or negatively affect their operations.

\subsubsection{Roles \& Structures}\label{sec:roles}
\begin{figure}
    \centering
    \includegraphics[width=\linewidth]{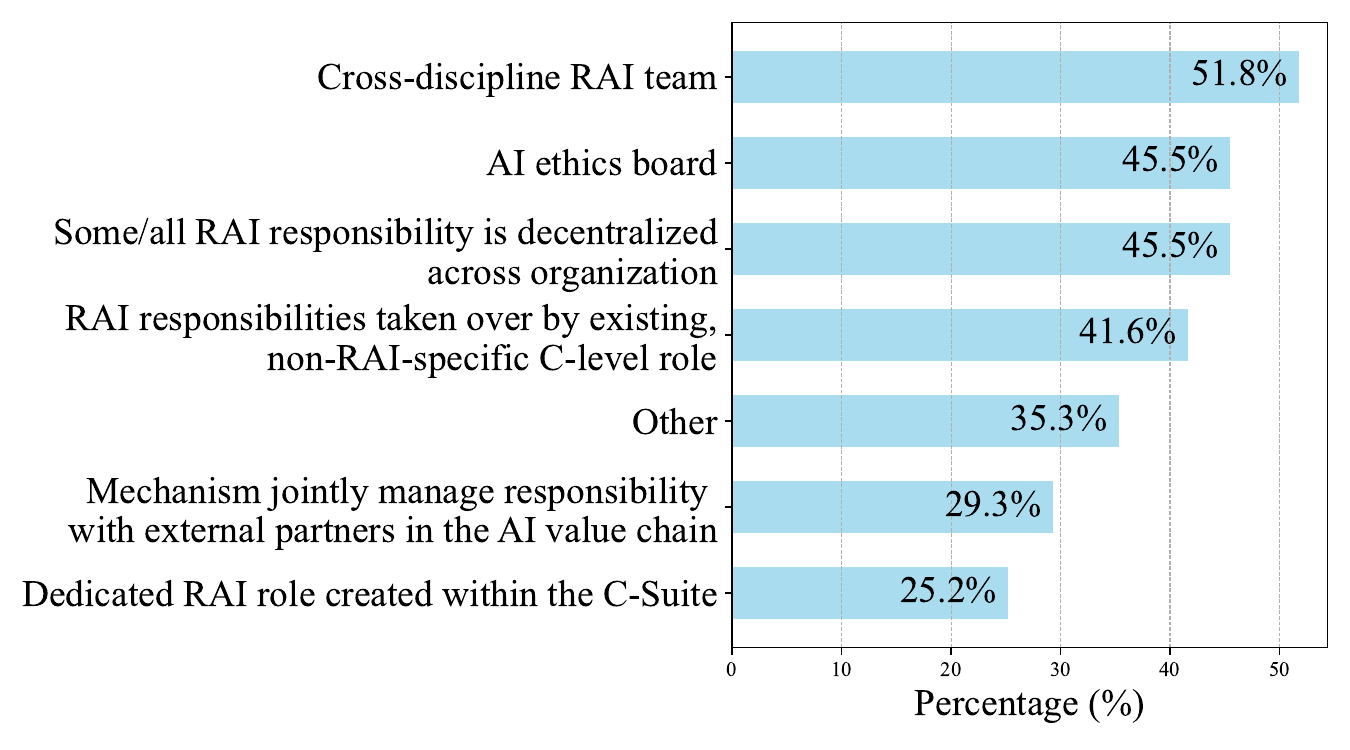}
    \caption{Reported RAI roles and structures within organizations.}
    \label{fig:roles}
\end{figure}

The majority of organizations~(51.8\%) indicated that they have a cross-discipline RAI team, followed by 45.5\% of respondents saying that some or all of their RAI responsibility is decentralized across business units and functions (Fig.~\ref{fig:roles}). 45.5\% indicated that they have an AI ethics board (or equivalent), while only 25.2\% indicated having a dedicated C-level RAI role.

\begin{figure}
    \centering
    \includegraphics[width=\linewidth]{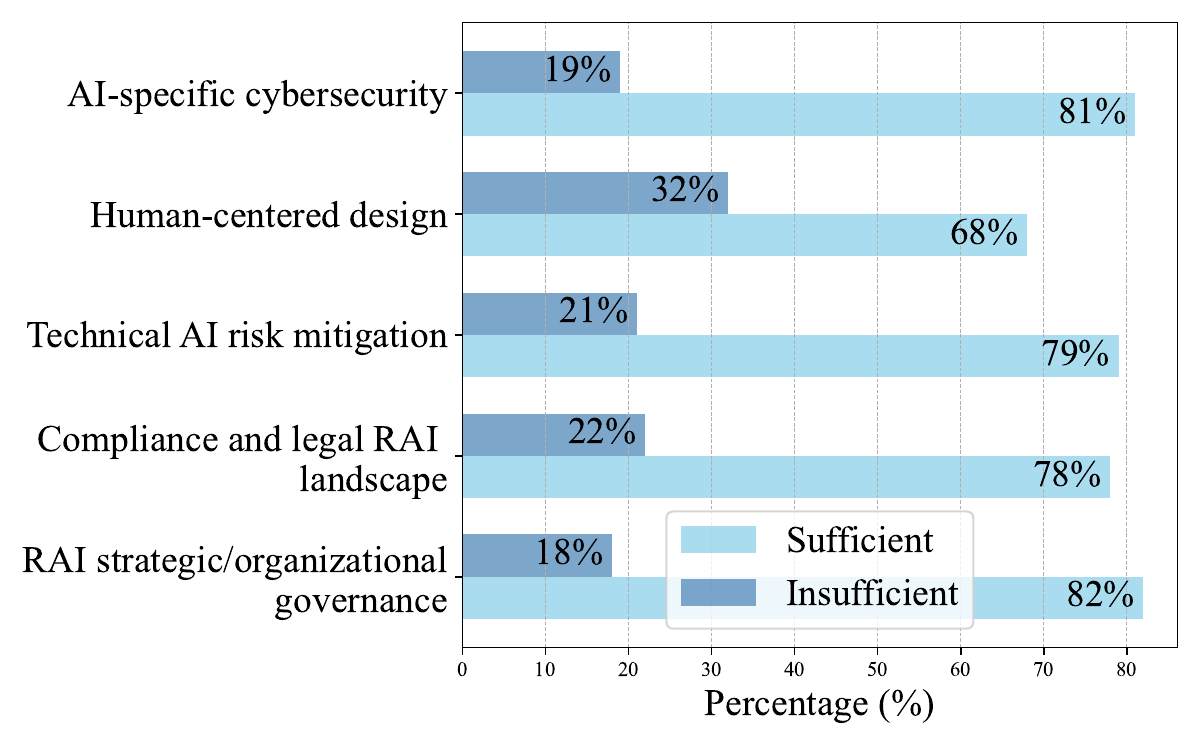}
    \caption{Indicated availability of sufficient talent with specific RAI skills.}
    \label{fig:talent}
\end{figure} 

\subsubsection{Sufficient Talent for Key RAI Skills}\label{sec:talent}
Companies were further asked to indicate if they have sufficient talent currently available to them for key RAI skills. The results present a relatively uniform picture, with about 20\% of companies indicating that they have insufficient people with AI-specific cybersecurity, technical AI risk mitigation, legal \& regulatory RAI, and strategic or organizational RAI governance skills. The exception is human-centered design, where 32\% of organizations indicated that they do not have enough talent with this skill set to meet current demands.

\subsubsection{Barriers to Generative AI Adoption}\label{sec:barriers}
When it comes to barriers to adopting generative AI, organizations prioritize harms~(43\%), data and privacy concerns~(39\%), and regulatory challenges~(34\%) over traditional hurdles such as talent acquisition~(31\%), budget constraints~(27\%), and scalability issues~(26\%). This shift highlights a growing awareness of the unique challenges posed by generative AI.

\begin{figure}
    \centering
    \includegraphics[width=\linewidth]{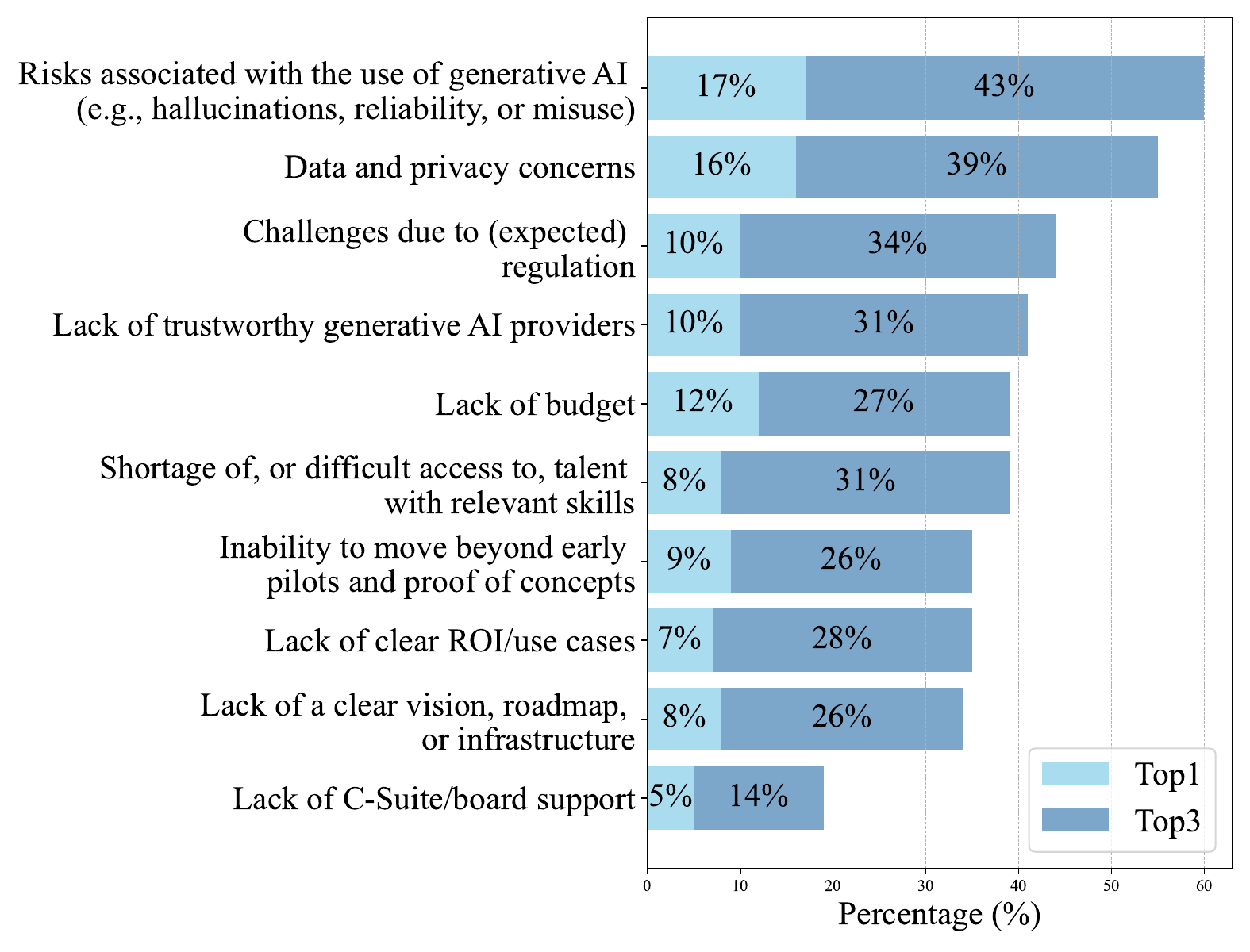}
    \caption{Barriers to the use and development of generative AI.}
    \label{fig:barriers}
\end{figure}

%[Add results of question 37A here on measures taken to mitigate risks from third-party generative AI models]

\section{Discussion}\label{sec:dis}
\begin{figure}
    \centering
    \includegraphics[width=\linewidth]{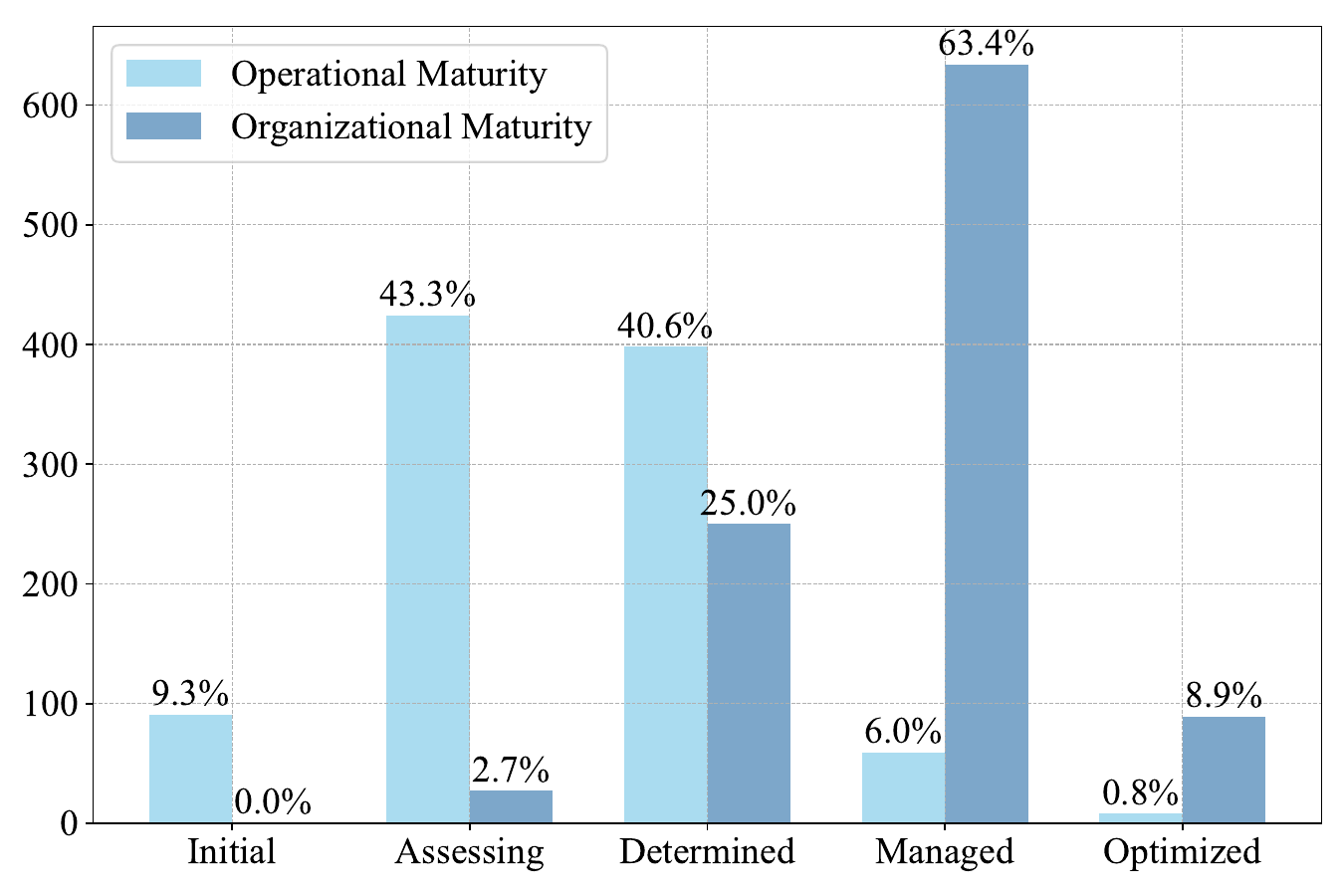}
    \caption{Comparison of organizational and operational maturity levels across all the organizations.}
    \label{fig:orgop}
\end{figure}

\textbf{Lack of Responsible AI Maturity.} 
No organization has reached the \textit{Optimized} stage in both organizational and operational dimensions. 9\% have achieved \textit{Optimized} organizational maturity, but only 0.8\% have reached operational maturity (Fig.~\ref{fig:orgop}), indicating a gap between planning and execution of RAI practices. This suggests formal RAI structures and policies exist, but implementation lags. The discrepancy between organizational and operational maturity poses a societal risk. Organizations might appear more prepared to handle AI responsibly than they actually are in practice. This could lead to a false sense of security among stakeholders and potentially inadequate safeguards against AI-related risks. The operational distribution's left skew indicates widespread difficulties implementing RAI practices, reflecting challenges in translating RAI principles and frameworks into operations. A statistically significant negative association exists between the number of relevant risks and operational maturity score ($p<0.05$, App.~\ref{app:regression_results}). There's also a negative correlation between regulatory exposure and both maturity types. This suggests that increased risk and regulatory exposure complicates system-level mitigation efforts, a potential indicator for the complexity of RAI implementation efforts.\\

\begin{figure}
    \centering
    \includegraphics[width=\linewidth]{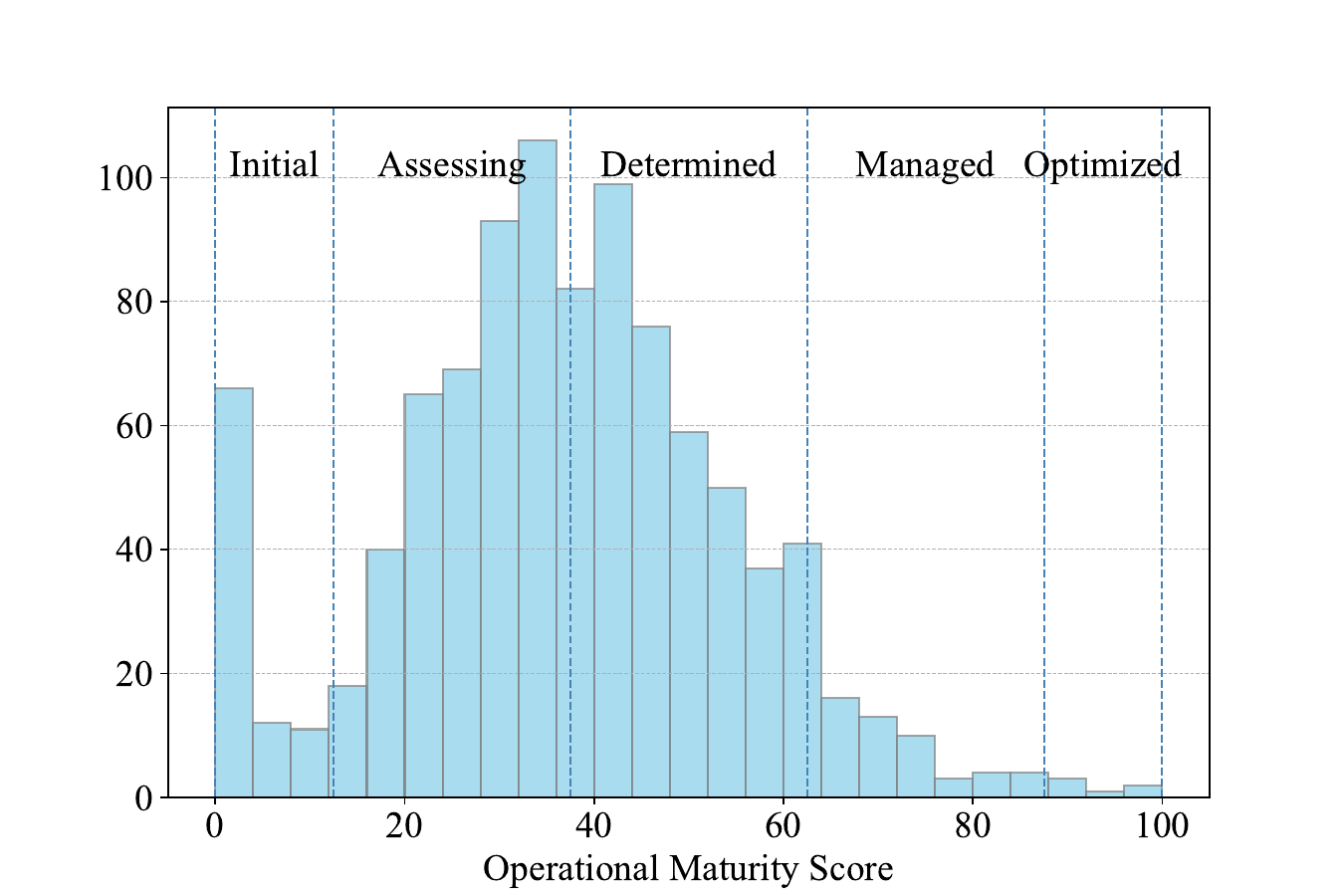}
    \caption{Operational RAI maturity distribution.}
    \label{fig:op_dist}
\end{figure}

\textbf{Responsibility Uncertainties.} The adoption of generative AI is significantly impeded by unresolved risks and perceived regulatory challenges. Yet, companies developing foundational models are seen as responsible for mitigating associated risks, rather than the organizations utilizing these models. This perspective is reflected in the operational maturity distribution, where a significant portion of entities cluster in the \textit{Initial} and \textit{Assessing} categories, potentially indicating a hesitancy among organizations to fully adopt mitigation measures due to unclear risk management responsibilities. Existing regulations, such as the EU AI Act and consumer protection laws, should be examined to clarify these responsibilities and ensure alignment with regulatory expectations. This regulatory clarity could potentially shift the entire distribution rightward, as organizations would have clearer guidelines for RAI development and deployment.\\

\textbf{Humans at the center of RAI.} 
Companies lack sufficient talent for effective RAI activities, especially in human-centered design, with human interaction being the least developed dimension in operational RAI maturity. 72\% of companies further indicated employees and end-users are critical in identifying and mitigating risks like hallucinations, cybersecurity threats, and IP/data breaches. However, ongoing RAI training for employees is limited, leaving them unprepared for evolving risks, potentially increasing public safety risks. The human-centered design skills shortage could result in AI systems misaligned with human values and ethics, potentially harming vulnerable populations or reinforcing societal biases. This might erode trust in AI systems, slowing adoption and the realization of societal benefits. The gap between the recognized importance of employee involvement and the lack of ongoing training may create challenges for regulators. It could be difficult to enforce RAI practices when the workforce lacks the necessary skills and knowledge to do so.

\section{Limitations}\label{sec:limit}
The study only included companies with annual global revenue exceeding 499 million USD, potentially excluding smaller but relevant entities. The survey provides a sampled perspective rather than complete data, and responses are inherently subjective, with potential cultural influences affecting uniformity. Surveys tend to be subject to a positive response bias compared to other means of data collection. Design decisions, such as not providing a `Not Applicable' option for some questions, could have impacted responses, too. The survey methodology is subject to biases, including social-desirability bias, convenience sampling, non-response bias, and self-selection bias, leading to a sample that may not accurately represent the target population. Techniques such as anonymity, pre-testing, indirect questions, response timing, and reverse-coding were employed to minimize these effects. These limitations underscore the need for cautious extrapolation of the findings to the broader population.

\section{Conclusion}\label{sec:conclusion}
Despite progress towards RAI maturity, significant gaps in organizational and operational implementation persist, potentially exposing society to unmitigated AI risks by organizations developing, deploying, and using AI. These gaps could lead to inadequate safeguards and AI systems misaligned with broader societal values. The difficulty in translating ethical principles into practice, especially in human-centered design and interaction, coupled with limited employee training on evolving AI risks, presents critical societal challenges. These issues could lead to increased public safety risks and eroded trust in AI systems, potentially slowing AI adoption and its societal benefits. Addressing these challenges requires collaborative efforts from industry, academia, and policymakers to bridge the gap between RAI planning and execution, focusing on human-centered approaches and comprehensive risk mitigation strategies to ensure AI advancement aligns with societal values.

%\section{Acknowledgments}

\bibliography{aaai25}

\section{Checklist}

This paper:

\begin{itemize}
\item Includes a conceptual outline and/or pseudocode description of AI methods introduced (YES, Sec.~\ref{sec:raimm} and Sec.~\ref{sec:method})
\item Clearly delineates statements that are opinions, hypothesis, and speculation from objective facts and results (YES)
\item Provides well marked pedagogical references for less-familiare readers to gain background necessary to replicate the paper (YES)
\item Does this paper make theoretical contributions? If yes, please complete the list below. (NO) 
\end{itemize}

\begin{itemize}
\item All assumptions and restrictions are stated clearly and formally. (NA)
\item All novel claims are stated formally (e.g., in theorem statements). (NA)
\item Proofs of all novel claims are included. (NA)
\item Proof sketches or intuitions are given for complex and/or novel results. (NA)
\item Appropriate citations to theoretical tools used are given. (NA)
\item All theoretical claims are demonstrated empirically to hold. (NA)
\item All experimental code used to eliminate or disprove claims is included. (NA)
\item Does this paper rely on one or more datasets? If yes, please complete the list below (NO)
\end{itemize}

\begin{itemize}
\item A motivation is given for why the experiments are conducted on the selected datasets (NA)
\item All novel datasets introduced in this paper are included in a data appendix. (NA)
\item All novel datasets introduced in this paper will be made publicly available upon publication of the paper with a license that allows free usage for research purposes. (NA)
\item All datasets drawn from the existing literature (potentially including authors’ own previously published work) are accompanied by appropriate citations. (NA)
\item All datasets drawn from the existing literature (potentially including authors’ own previously published work) are publicly available. (NA)
\item All datasets that are not publicly available are described in detail, with explanation why publicly available alternatives are not scientifically satisficing. (NA)
\item Does this paper include computational experiments? If yes, please complete the list below. (NO) 
\end{itemize}

\begin{itemize}
\item Any code required for pre-processing data is included in the appendix. (NA)
\item All source code required for conducting and analyzing the experiments is included in a code appendix. (NA)
\item All source code required for conducting and analyzing the experiments will be made publicly available upon publication of the paper with a license that allows free usage for research purposes. (NA)
\item All source code implementing new methods have comments detailing the implementation, with references to the paper where each step comes from (NA)
\item If an algorithm depends on randomness, then the method used for setting seeds is described in a way sufficient to allow replication of results. (NA)
\item This paper specifies the computing infrastructure used for running experiments (hardware and software), including GPU/CPU models; amount of memory; operating system; names and versions of relevant software libraries and frameworks. (NA)
\item This paper formally describes evaluation metrics used and explains the motivation for choosing these metrics. (NA)
\item This paper states the number of algorithm runs used to compute each reported result. (NA)
Analysis of experiments goes beyond single-dimensional summaries of performance (e.g., average; median) to include measures of variation, confidence, or other distributional information. (NA)
\item The significance of any improvement or decrease in performance is judged using appropriate statistical tests (e.g., Wilcoxon signed-rank). (NA)
\item This paper lists all final (hyper-)parameters used for each model/algorithm in the paper’s experiments. (NA)
\item This paper states the number and range of values tried per (hyper-) parameter during development of the paper, along with the criterion used for selecting the final parameter setting. (NA)
\end{itemize}

\section{Appendix}

\subsection{Maturity Levels for Subcomponents}\label{app:subcomponent_maturitylvls}

Tab.~\ref{tab:subcomponents_maturity1} and Tab. ~\ref{tab:subcomponents_maturity2} below describe the maturity levels that we developed and that were used as part of the survey.

\begin{table*}[]
    \small
    \centering
    \begin{tabular}{>{\raggedright\arraybackslash}p{2.5cm} >{\raggedright\arraybackslash}p{2.5cm} >{\raggedright\arraybackslash}p{2.5cm} >{\raggedright\arraybackslash}p{2.5cm} >
    {\raggedright\arraybackslash}p{2.5cm} >
    {\raggedright\arraybackslash}p{2.5cm}}
        \toprule
        \textbf{Sub Component} & \textbf{Level 1} & \textbf{Level 2} & \textbf{Level 3} & \textbf{Level 4} & \textbf{Level 5}\\
        \midrule
        Governance Operating Model & None/Use of existing governance operating model with no/limited RAI focus & RAI-specific governance framework designed, following gap analysis of existing governance model against AI risks & RAI-specific framework translated into organization-wide operating model with RAI-specific roles and responsibilities & RAI-specific governance operating model fully operationalized across the organization, with clear RAI roles and responsibilities, processes, reporting, tracking, KPIs and incentives for front-
line personnel & Fully-operationalized RAI-specific governance model that is extended beyond the organization
to incorporate partners in the supply chain/ecosystem to ensure overarching responsibility with
respect to their AI models/systems\\
        \midrule
        Risk Management & No risk management framework in place. & Use of the existing, non-AI-specific risk management framework. & Performed a gap analysis of the existing risk management processes completed against all major AI risks, but no defined RAI risk management strategy and/or framework yet. & Defined or rolling out an RAI risk management strategy and an organization-wide RAI risk
management framework. & Fully operationalized RAI risk management framework across the organization and dedicated resources to ensure the approach stays up to date in light of advances in AI, new RAI risk management best practices and policy developments.\\ 
        \midrule
        Risk Identification & No approach to identifying AI risks beyond performance metrics (e.g., accuracy)
and no awareness of the risks. & Awareness of the applicable risk(s) but no structured approach to identifying them. & Defined or rolling out a structured approach to risk identification, based on an analysis of
the intended purpose of the AI model/system, foreseeable/known forms of misuses and
potentially affected stakeholders. & Fully operationalized structured approach to risk identification. & Fully operationalized  structured approach to risk identification, with dedicated
resources in place to proactively anticipate new risks.\\
        \midrule
        Risk Mitigation & No processes in place to mitigate identified RAI risks. & Ad-hoc mitigation techniques, typically led by data scientists/RAI practitioners at project level. & Structured processes to integrate some mitigation techniques into existing AI development processes. & Structured processes to integrate a comprehensive set of mitigation techniques into existing AI development
processes with clear stage gates, testing/evaluation procedures and multi-disciplinary reviews. & Structured processes to integrate a comprehensive set of mitigation processes and techniques into existing AI
development processes with clear stage gates, testing/evaluation procedures, multi-disciplinary
reviews, and proactive research on new advances in risk mitigation. \\
        \bottomrule
    \end{tabular}
    \caption{Maturity levels for organizational RAI subcomponents.}
    \label{tab:subcomponents_maturity1}
\end{table*} 

\begin{table*}[h!]
    \small
    \centering
    \begin{tabular}{>{\raggedright\arraybackslash}p{2.5cm} >{\raggedright\arraybackslash}p{2.5cm} >{\raggedright\arraybackslash}p{2.5cm} >{\raggedright\arraybackslash}p{2.5cm} >
    {\raggedright\arraybackslash}p{2.5cm} >
    {\raggedright\arraybackslash}p{2.5cm}}
        \toprule
        \textbf{Sub Component} & \textbf{Level 1} & \textbf{Level 2} & \textbf{Level 3} & \textbf{Level 4} & \textbf{Level 5}\\
        \midrule
        Monitoring \& Control & No specific monitoring and control process in place & Some general (non-RAI specific) monitoring by MLOps and/or similar teams within the
organization & Centralized plan for RAI monitoring and control defined/rolling out to monitor and identify risks
in production, with defined cross-domain responsibilities, roles, and processes & RAI monitoring and control system fully operationalized to monitor and identify risks in
production, with defined cross-domain responsibilities, roles, and processes & Monitoring and control systems fully operationalized and optimized to proactively and
continuously identify new RAI risks\\
        \midrule
        Cybersecurity & No awareness of any AI-related cybersecurity risks and no related
measures in place. & Awareness of AI-related cybersecurity risks but no related
cybersecurity measures in place. & Use of the existing cybersecurity policies in an ad-hoc manner to address AI-related incidents. & Clearly outlined AI-specific cybersecurity incident response plan for containing,
investigating, and rectifying AI-related events. & Clear AI-specific cybersecurity incident response plan with vulnerability management
measures for model audits, adversarial testing, etc., within the AI risk-management framework. \\
        \midrule
        Sponsorship & No active/explicit involvement &  RAI acknowledged, but not yet part of broader organization strategy & Advocates on RAI, and is involved in some RAI efforts, e.g., initiating RAI principles & Takes full responsibility on RAI and leads efforts, e.g., setting up RAI teams, overseeing
implementation of organization-wide RAI policies and specifying RAI-related C-level KPIs & Takes full responsibility on RAI and leads efforts, while also working with ecosystem of
peers/experts/regulators to evolve RAI initiatives \\
        \midrule
        Training & No official RAI training currently available & Ad-hoc RAI training in siloes/by project & Guidance and high-level RAI training provided for those holding RAI responsibilities, including end
users, business owners, data scientists, etc. & RAI training is part of all AI-related employee training. One-time role-specific RAI training
provided to all employees with RAI responsibilities, including end users, business owners, data
scientists, etc. & RAI training is part of all AI-related employee training. Ongoing role-specific RAI training
provided to all employees with RAI responsibilities including end users, business owners, data
scientists, etc. \\
        \bottomrule
    \end{tabular}
    \caption{Maturity levels for organizational RAI subcomponents (continued).}
    \label{tab:subcomponents_maturity2}
\end{table*} 

\newpage

\subsection{Survey Questions}\label{app:survey_questions}

This section contains the questions that respondents were asked as part of the survey. If the answer options are enumerated, respondents could only select one answer option. The enumeration is missing in case of questions where respondents could select multiple answers. Information in brackets [ ] describes the survey logic, e.g., what answers led to survey terminations (e.g., due to an exclusion based on the qualifier questions) or which questions where only asked if a precondition was fulfilled. For some question, a scale was indicated and respondents had to select, for each subquestion listed, the answer selecting from the defined scale.\\

\begin{survey}

\begin{question}[Which of the following most closely matches your current job function?]
\begin{choices}
\item Board Member
\item Chief Executive Officer (including business units \& geographic/market CEOs)
\item Chief Compliance Officer (or equivalent)
\item Chief Information Security/Cyber Security Officer (or equivalent)
\item Chief Risk Officer
\item Chief Digital Officer
\item Chief Analytics/AI Officer/Chief Data and Analytics Officer/Chief Data Officer/ or equivalent
\item Chief Information Officer/Chief Technology Officer
\item SVP/VP/Director of AI/ML Engineering or equivalent (reporting to any of the above titles)
\item Chief Data Scientist
\item Chief Legal Counsel (or equivalent)
\item Other [Terminate]
\end{choices}
\end{question}

\begin{question}[In which country is your organization headquartered?]
\begin{choices}
\item United Arab Emirates
\item Argentina
\item Australia
\item Brazil
\item Canada
\item China
\item Germany
\item Denmark
\item Spain
\item Finland
\item France
\item United Kingdom
\item India
\item Italy
\item Japan
\item Mexico
\item Norway
\item Saudi Arabia
\item Sweden
\item Singapore
\item United States
\item South Africa
\item Other [Terminate]
\end{choices}
\end{question}

\begin{question}[In what countries does your organization currently produce (i.e., design, develop, implement), use or sell AI? Select all that apply.]
\begin{choices}
\item United Arab Emirates
\item Argentina
\item Australia
\item Brazil
\item Canada
\item China
\item Germany
\item Denmark
\item Spain
\item Finland
\item France
\item United Kingdom
\item India
\item Italy
\item Japan
\item Mexico
\item Norway
\item Saudi Arabia
\item Sweden
\item Singapore
\item United States
\item South Africa
\item Other [Terminate IF ONLY]
\item None; we don't produce, use, or sell AI [Exclusive, Terminate]
\end{choices}
\end{question}

\begin{question}[Which industry does your organization mainly operate in?]
\begin{choices}
\item Aerospace \& Defense
\item Automotive
\item Banking
\item Capital Markets
\item Chemicals
\item Telecommunications, Media \& Entertainment
\item Consumer Goods \& Services
\item Energy
\item Healthcare
\item High Tech
\item Industrial Equipment
\item Insurance
\item Life Sciences
\item Natural Resources
\item Public Services
\item Retail
\item Software \& Platforms
\item Travel \& Transport
\item Utilities
\item Other [Terminate IF ONLY]
\end{choices}
\end{question}

\begin{question}[What was the total annual GLOBAL revenue of your organization in the latest financial year (including all regions and business units)?]
\begin{choices}
\item Less than \$1 million [Terminate]
\item \$1 to \$49 million [Terminate]
\item \$50 to \$99 million [Terminate]
\item \$100 to \$499 million [Terminate]
\item \$500 to \$999 million
\item \$1 to \$4.9 billion
\item \$5 to \$9.9 billion
\item \$10 to \$19.9 billion
\item \$20 to \$49.9 billion
\item \$50 billion or more
\end{choices}
\end{question}

\begin{question}[Which of the following statements best describes your role regarding the development and implementation of, and/or compliance with responsible AI (RAI) related strategies and priorities within your organization?]
\begin{choices}
\item We currently \underline{do not have a specific RAI strategy} or plan in place
\item Informed only after the strategy is finalized with \underline{responsibilities to execute} on the plan
\item Acting in the capacity of an \underline{advisor} to the C-suite and the board in these aspects, but do not have any influence on the final strategy
\item \underline{Fully involved} and working in collaboration with the C-suite and the board, business, and functions in setting the vision, performance criteria, KPIs, etc., but \underline{do not share} direct ownership and responsibility
\item \underline{Fully involved} and working in collaboration with the C-suite and the board, business, and functions in setting the vision, performance criteria, KPIs, etc., and \underline{share} direct ownership and responsibility
\item \underline{Not involved} or have any visibility to the RAI decision-making process [Terminate]
\end{choices}
\end{question}

\begin{question}[How many AI models/systems are you currently using/developing or planning to use/develop in the next two years in your organization?]
\begin{scale}{
\item 0
\item 1
\item 2 to 5
\item 6 to 10
\item 11 to 20
\item 21 or more
}
\item Currently using/developing
\item Using/developing in next 2 years [Terminate IF = 1]
\end{scale}
\end{question}

\begin{question}[Which of the following statements best describes your current AI adoption and implementation strategy?]
\begin{choices}
\item We do not use AI and an AI adoption strategy is irrelevant to us. [Terminate]
\item Organization has an \underline{ad-hoc AI strategy} with nascent data practices, model development/procurement processes, etc.
\item Organization has \underline{defined/is rolling out an AI strategy}, covering key areas like governance, roles, data, model development/procurement, tooling, MLOps, talent, partnerships, etc.
\item Organization-wide \underline{AI strategy is fully operationalized}, with a sufficiently responsible data strategy, centrally managed tools, and basic AI procurement and talent strategy.
\item Organization \underline{has processes in place to} continuously evolve and optimize all components of its operationalized AI strategy.
\end{choices}
\end{question}

\begin{question}[Which of the following risks are relevant to your current and future AI models/systems? Select all that apply.]
\begin{checkboxes}
\item Diversity \& Non-discrimination risks (e.g., fairness concerns, toxicity, discrimination and stereotype reproduction)
\item Privacy \& Data Governance risks (e.g., data leakage, unauthorized usage of data, etc.)
\item Reliability risks (e.g., output errors, hallucinations, model failure)
\item Security risks (e.g., cybersecurity incidents)
\item Human Interaction risks (e.g., misuse by users for the generation of deepfakes or misinformation, overreliance of users/employee on AI models/systems, or physical/mental harm due to model/system usage)
\item Transparency risks (e.g., inexplicable model decisions)
\item Societal risks (e.g., threats due to (semi-) autonomous decisions, threats to political stability, national security concerns)
\item Environmental risks (e.g., high carbon footprint of model training, inference, hardware)
\item Client/Customer risks (e.g., loss of trust, market share, customer satisfaction)
\item Brand/Reputational risks (e.g., damage caused to brand by AI-related incident)
\item Accountability risks (e.g., IP/copyright violations)
\item Compliance and Lawfulness risks (e.g., incompliance with AI regulations or related laws)
\item Organizational/Business risks (e.g., lack of AI ROI, AI-related financial loss)
\item Other (please specify)
\item None [Exclusive]
\end{checkboxes}
\end{question}

\begin{question}[Do you expect your organization to be subject to any specific regulation of AI/legal liabilities over the next 5 years, based on your current AI strategy, adoption, and geographical locations?]
\begin{choices}
\item Yes, we are already subject to such regulation of AI/legislations
\item Yes, within 3 years
\item Yes, within 5 years
\item No [Terminate IF q9 = 15]
\item Unsure
\end{choices}
\end{question}

\begin{question}[Please list the (types of) regulation of AI/legal liabilities that you feel apply/will apply to your organization (e.g., EU AI Act, New York AI Bias Law, EU Liability Directive, future national-level AI regulation, future AI-cybersecurity laws).]
\emph{[Free text, Optional]}
\end{question}

\medskip

\begin{question}[Please list the (types of) regulation of AI/legal liabilities that you feel could apply to your organization (e.g., EU AI Act, New York AI Bias Law, EU Liability Directive, future national-level AI regulation, future AI-cybersecurity laws).]
\emph{[Free text, Optional]}
\end{question}

\medskip

\begin{question}[Looking first at Generative AI, what is your organization's current and future (i.e., within two years) adoption strategy?]

\textbf{Current. Select all that apply.}
\begin{checkboxes}
\item Build and Use: \underline{Develop} your own (proprietary) generative AI models/systems from the ground up, to \underline{use internally} in production for AI-enabled processes and/or AI-enabled customer products and services.
\item Build and Sell: \underline{Develop} your own (proprietary) generative AI models/systems from the ground up, \underline{to make available commercially} to other \underline{businesses/organizations}.
\item Build and Open-Source: \underline{Develop} your own generative AI models/systems from the ground up, \underline{to open source} to other \underline{businesses/organizations}.
\item Resell: Partner on or gain access to \underline{third-party} generative AI models/systems to \underline{resell} (modified or unmodified) to other \underline{businesses/organizations}.
\item Use Third Party Models/Systems: \underline{Use third-party} generative AI models/systems (modified or unmodified) \underline{internally} in production for AI-enabled processes and/or AI-enabled customer products and services.
\item Other
\item We have no plans to use/develop/sell/open-source generative AI models/systems [Exclusive]
\end{checkboxes}

\textbf{Next 2 years. Select all that apply.}
\begin{checkboxes}
\item Build and Use: \underline{Develop} your own (proprietary) generative AI models/systems from the ground up, to \underline{use internally} in production for AI-enabled processes and/or AI-enabled customer products and services.
\item Build and Sell: \underline{Develop} your own (proprietary) generative AI models/systems from the ground up, \underline{to make available commercially} to other \underline{businesses/organizations}.
\item Build and Open-Source: \underline{Develop} your own generative AI models/systems from the ground up, \underline{to open source} to other \underline{businesses/organizations}.
\item Resell: Partner on or gain access to \underline{third-party} generative AI models/systems to \underline{resell} (modified or unmodified) to other \underline{businesses/organizations}.
\item Use Third Party Models/Systems: \underline{Use third-party} generative AI models/systems (modified or unmodified) \underline{internally} in production for AI-enabled processes and/or AI-enabled customer products and services.
\item Other
\item We have no plans to use/develop/sell/open-source generative AI models/systems [Exclusive]
\end{checkboxes}
\end{question}

\begin{question}[Now, looking at the wider AI space (\underline{excluding} Generative AI), what is your organization's current and future (i.e., within two years) AI adoption strategy?]

\textbf{Current. Select all that apply.}
\begin{checkboxes}
\item Build and Use: \underline{Develop} your own (proprietary) AI models/systems from the ground up, to \underline{use internally} in production for AI-enabled processes and/or AI-enabled customer products and services.
\item Build and Sell: \underline{Develop} your own (proprietary) AI models/systems from the ground up, \underline{to make available commercially} to other \underline{businesses/organizations}.
\item Build and Open-Source: \underline{Develop} your own AI models/systems from the ground up, \underline{to open source} to other \underline{businesses/organizations}.
\item Resell: Partner on or gain access to \underline{third-party} AI models/systems to \underline{resell} (modified or unmodified) to other \underline{businesses/organizations}.
\item Use Third Party Models/Systems: \underline{Use third-party} AI models/systems (modified or unmodified) \underline{internally} in production for AI-enabled processes and/or AI-enabled customer products and services.
\item Other
\item We have no plans to use/develop/sell/open-source AI models/systems [Exclusive]
\end{checkboxes}

\textbf{Next 2 years. Select all that apply.}
\begin{checkboxes}
\item Build and Use: \underline{Develop} your own (proprietary) AI models/systems from the ground up, to \underline{use internally} in production for AI-enabled processes and/or AI-enabled customer products and services.
\item Build and Sell: \underline{Develop} your own (proprietary) AI models/systems from the ground up, \underline{to make available commercially} to other \underline{businesses/organizations}.
\item Build and Open-Source: \underline{Develop} your own AI models/systems from the ground up, \underline{to open source} to other \underline{businesses/organizations}.
\item Resell: Partner on or gain access to \underline{third-party} AI models/systems to \underline{resell} (modified or unmodified) to other \underline{businesses/organizations}.
\item Use Third Party Models/Systems: \underline{Use third-party} AI models/systems (modified or unmodified) \underline{internally} in production for AI-enabled processes and/or AI-enabled customer products and services.
\item Other
\item We have no plans to use/develop/sell/open-source AI models/systems [Exclusive]
\end{checkboxes}
\end{question}

\begin{question}[What percentage (approximately) of your AI budget are you spending/intending to spend in the next years on RAI measures?]
\begin{scale}{
\item 0\%
\item 1 to 5\%
\item 6 to 10\%
\item 11 to 20\%
\item 21 to 30\%
\item 31 to 40\%
\item 41\% or more
\item Unsure
}
\item Current
\item Next 2 years
\end{scale}
\end{question}

\begin{question}[Which statement best describes the extent to which the CEO and/or Board of Directors currently sponsors RAI adoption in your organization and is responsible for it?]
\begin{choices}
\item No active/explicit involvement
\item RAI acknowledged, but not yet part of broader organization strategy
\item Advocates on RAI, and is involved in some RAI efforts, e.g., initiating RAI principles
\item Takes full responsibility on RAI and leads efforts, e.g., setting up RAI teams, overseeing implementation of organization-wide RAI policies and specifying RAI-related C-level KPIs
\item Takes full responsibility on RAI and leads efforts, while also working with ecosystem of peers/experts/regulators to evolve RAI initiatives
\end{choices}
\end{question}

\begin{question}[In your organization, RAI is viewed as... Select all that apply.]
\begin{checkboxes}
\item Slowing down innovation and time to market
\item Not critical for our current usage of AI
\item A regulatory compliance and legal issue
\item An additional cost of doing business that is necessary to incur
\item A core part in shaping/driving our overall AI strategy
\item A competitive advantage and value driver that can also unlock new markets
\item A way to improve our/clients'/users' brand reputation and AI trustworthiness
\item A differentiation in attracting/retaining key talent
\item A strategic tool in better ensuring AI-related revenue growth
\item A strategic tool in avoiding potential losses/brand damage due to fines, cybersecurity, etc.
\item A way to demonstrate social responsibility
\item A way to industrialize our AI processes and improve the performance of our AI models/systems, time to market, etc.
\item Necessary to ensure the safety and security of our AI models/systems
\item Other (please specify)
\end{checkboxes}
\end{question}

\begin{question}[Please select which organization-wide RAI principles, guidelines, or policies are in place in your organization. Select all that apply.]
\begin{checkboxes}
\item RAI principles
\item RAI guidelines
\item RAI policies
\item Only team-level RAI principles, guidelines or policies
\item Other
\item None [Exclusive]
\end{checkboxes}
\end{question}

\begin{question}[Which statement \underline{best} describes your current RAI-specific governance operating model?]
\begin{choices}
\item None/Use of existing governance operating model with no/limited RAI focus
\item RAI-specific \underline{governance framework designed}, following gap analysis of existing governance model against AI risks
\item RAI-specific framework \underline{translated into organization-wide operating model} with RAI-specific roles and responsibilities
\item RAI-specific governance operating model \underline{fully operationalized} across the organization, with clear RAI roles and responsibilities, processes, reporting, tracking, KPIs and incentives for front-line personnel
\item Fully-operationalized RAI-specific governance model that is extended beyond the organization to incorporate \underline{partners in supply chain/ecosystem} to ensure overarching responsibility with respect to our AI models/systems
\end{choices}
\end{question}

\begin{question}[Which of the following dedicated RAI roles and structures are in place in your organization (either individually, or in collaboration) to take responsibility for RAI efforts? Select all that apply.]
\begin{checkboxes}
\item Cross-functional Governance Committee/Board: Consisting of multiple C-suite members, created to define and implement company RAI policies, processes, etc.
\item Cross-discipline RAI team that collaborates to build/test approaches to help ensure that AI models/systems are designed and used responsibly
\item AI ethics board
\item Dedicated RAI role created within the C-Suite (e.g., Chief Responsible AI/AI Ethics Officer)
\item RAI responsibilities are centralized with an existing, non-RAI-specific C-level role (e.g., CDO, CTO, CLO, CCO)
\item Some/all RAI responsibility is decentralized across business units and functions
\item Mechanism in place to jointly manage "responsibility" with external partners in the AI value chain
\item Other (please specify)
\item None [Exclusive]
\end{checkboxes}
\end{question}

\begin{question}[Which statement best describes the current RAI risk management framework in place at your organization?]
\begin{choices}
\item We don't currently have a risk management framework.
\item We use our existing risk, non-AI-specific risk management framework.
\item We conducted a gap analysis of the existing risk management processes completed against all major AI risks, but have yet to define an RAI risk management strategy and/or framework.
\item We have defined/rolling out an RAI risk management strategy and an organization-wide RAI risk management framework.
\item We have a fully operationalized RAI risk management framework across the organization, leveraging new risk management practices, AI lifecycle checkpoints and responsible parties.
\item We have a fully operationalized RAI risk management framework across the organization and dedicate resources to ensure our approach stays up to date in light of advances in AI, new RAI risk management best practices and policy developments.
\end{choices}
\end{question}

\begin{question}[Which statement best describes the process by which you identify risks in the development and/or use of your AI models/systems?]
\begin{choices}
\item We currently have no approach to identify AI risks beyond performance metrics (e.g., accuracy) and are not aware of the risks.
\item We are aware of the applicable risk(s) but there is no structured approach to identify them.
\item We have defined/rolling out a structured approach to risk identification, based on an analysis of the intended purpose of the AI model/system, foreseeable/known forms of misuses and potentially affected stakeholders.
\item We have fully operationalized a structured approach to risk identification.
\item We have fully operationalized a structured approach to risk identification, with dedicated resources in place to proactively anticipate new risks.
\end{choices}
\end{question}

\begin{question}[Which statement best describes your organization's current RAI risk mitigation processes during development of your AI models/systems?]
\begin{choices}
\item No processes or techniques used to mitigate identified RAI risks.
\item Ad-hoc mitigation techniques, typically led by data scientists/RAI practitioners at project level.
\item Structured integration of some mitigation techniques into existing AI development processes.
\item Integration of a comprehensive set of mitigation techniques into existing AI development processes with clear stage gates, testing/evaluation procedures and multi-disciplinary reviews.
\item Integration of a comprehensive set of mitigation processes and techniques into existing AI development processes with clear stage gates, testing/evaluation procedures, multi-disciplinary reviews, and proactive research on new advances in risk mitigation.
\end{choices}
\end{question}

\begin{question}[Which statement best describes how your organization applies the following measures in your current procurement evaluation processes for third-party providers when acquiring, licensing, or using AI models/systems.]
\begin{scale}{
\item Not applied
\item Assess -- We are assessing this measure but haven't used it yet
\item Ad-Hoc -- We have used this measure on an ad-hoc basis with our models/systems
\item Rolling out -- We have defined requirements and are rolling this out for all our relevant models/systems
\item Fully operationalized -- We require this measure to be implemented for all our relevant AI models/systems
}
\item Key metrics for RAI dimensions like performance, reliability, and fairness
\item Provider's RAI risk identification and evaluation processes
\item Provider's documentation of model limitations, instructions for use, interfaces and tailored training
\item Demonstration of alignment of model/system with emerging regulation of AI in relevant jurisdictions
\item Service agreements that encompass redress channels and remediation of RAI-related risks
\item Contractual agreements to ensure all RAI-related responsibilities have been agreed upon and are fully documented
\item Checks to ensure that the data used to train models was legally obtained
\end{scale}
\end{question}

\begin{question}[Which statement best describes the RAI monitoring and control processes that your organization has in place after you've deployed an AI model/system or after you've given access to your model/system to an external party?]
\begin{choices}
\item No specific monitoring and control process in place
\item Some general (non-RAI specific) monitoring by MLOps and/or similar teams within the organization
\item Centralized plan for RAI monitoring and control defined/rolling out to monitor and identify risks in production, with defined cross-domain responsibilities, roles, and processes
\item RAI monitoring and control system fully operationalized to monitor and identify risks in production, with defined cross-domain responsibilities, roles, and processes
\item Monitoring and control systems fully operationalized and optimized to proactively and continuously identify new RAI risks
\end{choices}
\end{question}

\begin{question}[What support do you provide to purchasers/users of your AI models, systems, and services (if any) to help them meet their RAI priorities or obligations (e.g., with respect to fairness, transparency, or compliance)? Select all that apply.]
\begin{checkboxes}
\item Mechanisms to provide support to purchasers/users with select RAI priorities and obligations, on request
\item Comprehensive guidance and support for purchasers to meet their RAI priorities and obligations, as standard
\item We have built our product to anticipate and meet RAI priorities and obligations and provide additional support where necessary
\item Other
\item No specific RAI support [Exclusive]
\end{checkboxes}
\end{question}

\begin{question}[Which statement best describes how your organization adopts cybersecurity in your AI risk management practices?]
\begin{choices}
\item We are not aware of any AI-related cybersecurity risks and do not have in place any related measures.
\item We are aware of the AI-related cybersecurity risks but do not have in place any related cybersecurity measures.
\item We use our existing cybersecurity policies in an ad-hoc manner to address AI-related incidents.
\item We have a clearly outlined AI-specific cybersecurity incident response plan for containing, investigating, and rectifying AI-related events.
\item We have clear AI-specific cybersecurity incident response plan with vulnerability management measures for model audits, adversarial testing, etc., within our AI risk-management framework.
\end{choices}
\end{question}

\begin{question}[Which statement best describes how your organization applies the following measures to mitigate human interaction risks.]
\begin{scale}{
\item Not applied
\item Assess -- We are assessing this measure but haven't used it yet
\item Ad-Hoc -- We have used this measure on an ad-hoc basis with our models/systems
\item Rolling out -- We have defined requirements and are rolling this out for all our relevant models/systems
\item Fully operationalized -- We require this measure to be implemented for all our relevant AI models/systems
}
\item Implementation of technical safeguards on top of the base model (e.g., filters) to decrease non-compliant behavior or misuse potential
\item Use of human-centered design techniques to specifically improve user understanding e.g., interface design, etc.
\item Monitor usage patterns for anomalous activity that may indicate misuse or attempts to game the system
\item Case/Role-specific training of users, and provision of information about the limitations of the AI model/system
\item Use of codes of conduct or terms of service to restrict misuse
\item Testing for potential misuse of AI model/system during evaluation
\end{scale}
\end{question}

\begin{question}[Which statement best describes how your organization applies the following measures to ensure sufficient data quality, both during training and with respect to the input data used during deployment.]
\begin{scale}{
\item Not applied
\item Assess -- We are assessing this measure but haven't used it yet
\item Ad-Hoc -- We have used this measure on an ad-hoc basis with our models/systems
\item Rolling out -- We have defined requirements and are rolling this out for all our relevant models/systems
\item Fully operationalized -- We require this measure to be implemented for all our relevant AI models/systems
}
\item Data collection and preparation include assessment of the completeness, uniqueness, consistency, and accuracy of the data
\item Remediation plans for and documentation of datasets with shortcomings
\item Checks to ensure that the data is representative with respect to the demographic setting within which the final model/system is used
\item Regular data audits and updates to ensure the relevancy of the data
\item Checks to ensure that the data complies with all relevant laws and regulations and is used with consent where applicable
\item Process for dataset documentation and traceability throughout the AI lifecycle
\end{scale}
\end{question}

\begin{question}[Which statement best describes how your organization applies the following measures to increase AI model/system fairness.]
\begin{scale}{
\item Not applied
\item Assess -- We are assessing this measure but haven't used it yet
\item Ad-Hoc -- We have used this measure on an ad-hoc basis with our models/systems
\item Rolling out -- We have defined requirements and are rolling this out for all our relevant models/systems
\item Fully operationalized -- We require this measure to be implemented for all our relevant AI models/systems
}
\item Collection of representative data based on the anticipated user demographics
\item Making methodology and data sources accessible to third-parties (auditors/general public) for independent oversight
\item Involvement of diverse stakeholders in model development and/or review process
\item Assessment of performance across different demographic groups
\item Use of technical bias mitigation techniques during model development
\end{scale}
\end{question}

\begin{question}[Which statement best describes how your organization applies the following measures to measure and minimize AI model/system environmental footprint.]
\begin{scale}{
\item Not applied
\item Assess -- We are assessing this measure but haven't used it yet
\item Ad-Hoc -- We have used this measure on an ad-hoc basis with our models/systems
\item Rolling out -- We have defined requirements and are rolling this out for all our relevant models/systems
\item Fully operationalized -- We require this measure to be implemented for all our relevant AI models/systems
}
\item Measurement of environmental footprint of AI models/systems
\item Provision of carbon impact statement for AI models/systems
\item Technical measures to minimize environmental impact during AI development (e.g., by implementing power-efficient coding practices or using eco-friendly hardware)
\item Technical measures to minimize environmental impact during use/deployment (e.g., by using energy-efficient infrastructure or setting energy-efficient default hyperparameters)
\item Carbon reduction strategies at the organization level (e.g., carbon offsetting or use of renewable energy)
\end{scale}
\end{question}

\begin{question}[Which statement best describes how your organization applies the following measures to validate and ensure AI model/system reliability.]
\begin{scale}{
\item Not applied
\item Assess -- We are assessing this measure but haven't used it yet
\item Ad-Hoc -- We have used this measure on an ad-hoc basis with our models/systems
\item Rolling out -- We have defined requirements and are rolling this out for all our relevant models/systems
\item Fully operationalized -- We require this measure to be implemented for all our relevant AI models/systems
}
\item Mitigation measures for model errors and to handle low confidence outputs
\item Failover plans or other measures to ensure the system's/model's availability
\item Evaluation of models/systems for vulnerabilities or harmful behavior (i.e., red teaming)
\item Measures to prevent adversarial attacks
\item Confidence scoring for model outputs
\item Comprehensive test cases that cover a wide range of scenarios and metrics
\end{scale}
\end{question}

\begin{question}[Which statement best describes how your organization applies the following measures to increase model/system transparency.]
\begin{scale}{
\item Not applied
\item Assess -- We are assessing this measure but haven't used it yet
\item Ad-Hoc -- We have used this measure on an ad-hoc basis with our models/systems
\item Rolling out -- We have defined requirements and are rolling this out for all our relevant models/systems
\item Fully operationalized -- We require this measure to be implemented for all our relevant AI models/systems
}
\item Model explainability/interpretability tools (e.g., saliency maps) to elucidate model decisions
\item Documentation of the development process, detailing algorithm design choices, data sources, intended use cases, and limitations
\item Prioritization of simpler models where high interpretability is crucial, even if it sacrifices some performance
\item Training programs for stakeholders (incl. users) covering the intended use cases and limitations of model
\end{scale}
\end{question}

\begin{question}[Which statement best describes how your organization applies the following measures to increase model/system accountability.]
\begin{scale}{
\item Not applied
\item Assess -- We are assessing this measure but haven't used it yet
\item Ad-Hoc -- We have used this measure on an ad-hoc basis with our models/systems
\item Rolling out -- We have defined requirements and are rolling this out for all our relevant models/systems
\item Fully operationalized -- We require this measure to be implemented for all our relevant AI models/systems
}
\item Logging and traceability mechanisms of a model's/system's outputs
\item Redress mechanisms for stakeholders
\item Negative incidence tracking and reporting
\item Publication of model/system documentation detailing algorithm design choices, data sources, intended use cases, and limitations
\item Third-party audits of models/systems for RAI considerations
\item Version control of models/systems to track changes, updates, or modifications
\item Regular model/system reviews to ensure they're aligned with original objectives
\end{scale}
\end{question}

\begin{question}[Which statement best describes how your organization applies the following measures to improve model/system cybersecurity.]
\begin{scale}{
\item Not applied
\item Assess -- We are assessing this measure but haven't used it yet
\item Ad-Hoc -- We have used this measure on an ad-hoc basis with our models/systems
\item Rolling out -- We have defined requirements and are rolling this out for all our relevant models/systems
\item Fully operationalized -- We require this measure to be implemented for all our relevant AI models/systems
}
\item Basic cybersecurity hygiene practices (e.g., multi-factor authentication, access controls, and employee training)
\item Dedicated AI cybersecurity team and/or personnel trained specifically for AI-specific cybersecurity
\item Technical AI-specific cybersecurity checks and measures, e.g., adversarial testing, vulnerability assessments, and data security measures
\item Resources dedicated to research and monitoring of evolving AI-specific cybersecurity risks and integration in existing cybersecurity processes
\item Vetting and validation of cybersecurity measures of third-parties in the supply chain
\end{scale}
\end{question}

\begin{question}[Which statement best reflects the current level of RAI training offered across your organization?]
\begin{choices}
\item No official RAI training currently available
\item Ad hoc training in siloes/by project
\item Guidance and high-level training provided for those holding RAI responsibilities, including end users, business owners, data scientists, etc.
\item RAI training is part of all AI-related employee training. \underline{One-time} role-specific RAI training provided to all employees with RAI responsibilities, including end users, business owners, data scientists, etc.
\item RAI training is part of all AI-related employee training. \underline{Ongoing} role-specific RAI training provided to all employees with RAI responsibilities including end users, business owners, data scientists, etc.
\end{choices}
\end{question}

\begin{question}[From the following list, please select the areas where you feel your organization has sufficient talent/skills to conduct RAI activities effectively in the next 2 years.]
\begin{scale}{
\item Sufficient
\item Insufficient
}
\item RAI strategic/organizational governance
\item Compliance and legal RAI landscape
\item Technical AI risk mitigation
\item Human-centered design
\item AI-specific cybersecurity
\end{scale}
\end{question}

\begin{question}[Please indicate how much you agree with the following statements applicability to your organization in the next 5 years.]
\begin{scale}{
\item Strongly disagree
\item Disagree
\item Neither agree nor disagree
\item Agree
\item Strongly agree
\item Not sure
}
\item My organization will be subject to liability laws and directives due to our AI adoption.
\item My organization is or will be subject to cybersecurity laws due to our AI adoption.
\item My organization is or will be subject to data protection and privacy laws due to our AI adoption.
\item My organization is or will be subject to consumer safety \& protection laws due to our AI adoption.
\item My organization expects to be subject to future generative AI law (e.g., regarding truthfulness/misinformation) due to our AI adoption.
\item My organization contributes to RAI best practice and standardization efforts.
\end{scale}
\end{question}

\begin{question}[What are the greatest barriers to the use/development of generative AI by your organization? Rank the top three, where 1 = greatest barrier.]
\begin{enumerate}
\item Lack of a clear vision, roadmap, infrastructure, etc.
\item Lack of C-Suite/board support
\item Lack of clear ROI/use cases associated with of generative AI
\item Inability to move beyond early pilots and proof of concepts to scaling across the organization
\item Shortage of, or difficult access to, talent with generative AI skills
\item Data and privacy concerns
\item Risks associated with the use of generative AI (e.g., hallucination/factually incorrect output/transparency and explainability, security and bias or misuse related concerns etc.)
\item Lack of budget
\item Challenges due to current/future regulation of generative AI
\item Lack of trustworthy generative AI providers
\item Other (please specify)
\end{enumerate}
\end{question}

\begin{question}[Which of the following measures do you have in place to mitigate risks when using generative AI models from a third-party provider? Select all that apply.]
\begin{checkboxes}
\item Provider Selection: Selection of model/system provider that prioritizes and supports RAI considerations
\item Evaluation: RAI-risk-focused testing of models (e.g., human feedback, red teaming for toxic/offensive output, adversarial testing)
\item Infrastructure: Running generative AI models/systems on infrastructure with appropriate security and privacy protocols, e.g., to prevent data leakages or system breaches
\item Application: Implementation of additional technical safeguards on top of third-party model/system to decrease non-compliant behavior or misuse potential (e.g., addition of safety packages or filters to prevent toxic/offensive content, hallucinations, malicious prompts, prompt injections, or PII \& sensitive data leaks)
\item End-User: Any measure targeted at end-users, e.g., use of codes of conduct or terms of service and training to restrict misuse
\item Monitoring, Control \& Observability: Monitoring/analysis of prompts, responses and model behavior to mitigate issues and improve performance
\item Other
\item None [Exclusive]
\end{checkboxes}
\end{question}

\begin{question}[Which of the following measures do you have in place to mitigate risks when developing generative AI models/systems? Select all that apply.]
\begin{checkboxes}
\item Infrastructure: Running generative AI systems on infrastructure with appropriate security and privacy protocols, e.g., to prevent data leakage or system breaches
\item Model: Technical risk mitigation approaches at the model level, e.g., reinforcement learning from human feedback or fine-tuning
\item Evaluation: RAI-risk-focused testing of models (e.g., evaluator GAIs, red teaming for toxic/offensive output)
\item Application: Implementation of additional technical safeguards on top of base model to decrease non-compliant behavior or misuse potential (e.g., addition of safety packages and filters to prevent toxic/offensive content, hallucinations, malicious prompts, prompt injections, or PII \& sensitive data leaks)
\item End-User: Any measure targeted at end-users, e.g., use of codes of conduct or terms of service and training to restrict misuse
\item Post-Deployment monitoring, control \& observability: Monitoring/analysis of prompts, responses and model behavior to mitigate issues and improve performance
\item Other
\item None [Exclusive]
\end{checkboxes}
\end{question}

\begin{question}[Please indicate how much you agree with the following statements.]
\begin{scale}{
\item Strongly disagree
\item Disagree
\item Neither agree nor disagree
\item Agree
\item Strongly agree
\item Not sure
}
\item Companies that develop foundation models will be responsible for the mitigation of all associated risks, rather than organizations using these models/systems.
\item The new and evolving business risks associated with generative AI mean that RAI will be essential for my organization to unlock and maintain its value over time.
\item Employees/end-users will play a pivotal role in the identification and mitigation of risks (e.g., hallucinations or cybersecurity and IP/data breaches).
\item I believe that generative AI presents enough of a threat that globally agreed generative AI governance is required.
\item Using open-source generative AI models is safer than using proprietary generative AI models because there's added public scrutiny and we can make direct modifications to the model, if necessary.
\end{scale}
\end{question}

\end{survey}

\newpage

\subsection{Additional Results}\label{app:add_results}
\subsubsection{Regression Results}\label{app:regression_results}
Tab.~\ref{tab:ols_results_org_1} presents the OLS regression analysis results for organizational maturity. The results indicate that the diversity and non-discrimination risks~(Q9-1) and privacy \& data governance risks~(Q9-2) have negative associations with organizational maturity, with coefficients of $-2.78$ and $-2.27$, respectively, and are statistically significant. Also, environmental risk~(Q9-8) and brand \& reputational risk have positive associations with organizational maturity, with coefficients of $2.81$ and $2.17$, respectively and are statistically significant. Other risk factors, such as security risks (Q9-4) and human interaction risks (Q9-5), show no statistically significant relationship with organizational maturity.\\

\begin{table*}[] 
\centering \begin{tabular}{>{\raggedright\arraybackslash}m{2cm} >{\raggedright\arraybackslash}m{2cm} >{\raggedright\arraybackslash}m{2cm} >{\raggedright\arraybackslash}m{2cm} >{\raggedright\arraybackslash}m{2cm} >{\raggedright\arraybackslash}m{3cm}} \toprule \textbf{Variable} & \textbf{Coefficient} & \textbf{Standard Error} & \textbf{t-Statistic} & \textbf{p-Value} & \textbf{95\% Confidence Interval} \\ \midrule Constant & 69.2131 & 1.267 & 54.614 & 0.000 & [66.726, 71.700] \\ q9\_1 & -2.7819 & 0.955 & -2.912 & 0.004 & [-4.657, -0.907] \\ q9\_2 & -2.2698 & 0.919 & -2.469 & 0.014 & [-4.074, -0.465] \\ q9\_3 & 0.6507 & 0.927 & 0.702 & 0.483 & [-1.169, 2.471] \\ q9\_4 & 0.1571 & 0.911 & 0.172 & 0.863 & [-1.631, 1.945] \\ q9\_5 & 0.3374 & 0.966 & 0.349 & 0.727 & [-1.558, 2.232] \\ q9\_6 & 1.3797 & 0.983 & 1.403 & 0.161 & [-0.549, 3.309] \\ q9\_7 & 0.3361 & 0.937 & 0.359 & 0.720 & [-1.502, 2.174] \\ q9\_8 & 2.8151 & 1.005 & 2.800 & 0.005 & [0.842, 4.788] \\ q9\_9 & 0.4066 & 0.960 & 0.424 & 0.672 & [-1.477, 2.290] \\ q9\_10 & 2.1711 & 1.006 & 2.158 & 0.031 & [0.196, 4.146] \\ q9\_11 & -0.5122 & 1.027 & -0.499 & 0.618 & [-2.528, 1.504] \\ q9\_12 & 0.2019 & 1.030 & 0.196 & 0.845 & [-1.819, 2.222] \\ q9\_13 & 0.3036 & 1.383 & 0.220 & 0.826 & [-2.410, 3.017] \\ \bottomrule \end{tabular} \caption{OLS regression results for organizational maturity.} \label{tab:ols_results_org_1} \end{table*}

Tab.~\ref{tab:ols_results_org_2} shows the relationship between the total number of risks identified by organizations~(sum-Q1-Q13) and their organizational maturity. The coefficient for sum-Q1-Q13 is $0.1892$, with a p-value of $0.400$, indicating that the total number of identified risks does not have a statistically significant impact on organizational maturity. This suggests that simply identifying more risks does not necessarily correlate with higher or lower organizational maturity levels.\\

\begin{table*}[]
\centering
\begin{tabular}{>{\raggedright\arraybackslash}m{2cm} >{\raggedright\arraybackslash}m{2cm} >{\raggedright\arraybackslash}m{2cm} >{\raggedright\arraybackslash}m{2cm} >{\raggedright\arraybackslash}m{2cm} >{\raggedright\arraybackslash}m{3cm}}
\toprule
\textbf{Variable} & \textbf{Coefficient} & \textbf{Standard Error} & \textbf{t-Statistic} & \textbf{p-Value} & \textbf{95\% Confidence Interval} \\
\midrule
Constant & 69.1004 & 1.090 & 63.420 & 0.000 & [66.962, 71.239] \\
sum\_q1\_q13 & 0.1892 & 0.225 & 0.842 & 0.400 & [-0.252, 0.630] \\
\bottomrule
\end{tabular}
\caption{OLS regression analysis results for organizational maturity.}
\label{tab:ols_results_org_2}
\end{table*}

Tab.~\ref{tab:ols_results_org_3} presents the regression results for the impact of the organization's expectation to be subject to AI-specific regulations or legal liabilities~(Q10) on organizational maturity. The results reveal a negative association with a coefficient of $-2.1318$ and a p-value of $0.000$, indicating statistical significance.\\ 

\begin{table*}[]
\centering
\begin{tabular}{>{\raggedright\arraybackslash}m{2cm} >{\raggedright\arraybackslash}m{2cm} >{\raggedright\arraybackslash}m{2cm} >{\raggedright\arraybackslash}m{2cm} >{\raggedright\arraybackslash}m{2cm} >{\raggedright\arraybackslash}m{3cm}}
\toprule
\textbf{Variable} & \textbf{Coefficient} & \textbf{Standard Error} & \textbf{t-Statistic} & \textbf{p-Value} & \textbf{95\% Confidence Interval} \\
\midrule
Constant & 74.9482 & 0.998 & 75.102 & 0.000 & [72.990, 76.907] \\
q10 & -2.1318 & 0.386 & -5.527 & 0.000 & [-2.889, -1.375] \\
\bottomrule
\end{tabular}
\caption{OLS regression analysis results for organizational maturity.}
\label{tab:ols_results_org_3}
\end{table*}

Tab.~\ref{tab:ols_results_op_1} summarizes the regression analysis of various AI-related risk factors on operational maturity. The results show that diversity and non-discrimination risks~(Q9-1) have a positive and statistically significant impact on operational maturity with a coefficient of $2.8510$. Conversely, privacy \& data governance risks~(Q9-2) and accountability risks~(Q9-12) have negative associations, with coefficients of $-0.2123$ and $-3.9899$, respectively. This suggests that while addressing diversity issues may enhance operational maturity, privacy and accountability concerns could pose challenges to operational maturity.\\

\begin{table*}[]
\centering
\begin{tabular}{>{\raggedright\arraybackslash}m{2cm} >{\raggedright\arraybackslash}m{2cm} >{\raggedright\arraybackslash}m{2cm} >{\raggedright\arraybackslash}m{2cm} >{\raggedright\arraybackslash}m{2cm} >{\raggedright\arraybackslash}m{3cm}}
\toprule
\textbf{Variable} & \textbf{Coefficient} & \textbf{Standard Error} & \textbf{t-Statistic} & \textbf{p-Value} & \textbf{95\% Confidence Interval} \\
\midrule
Constant & 38.3211 & 1.726 & 22.202 & 0.000 & [34.934, 41.708] \\
q9\_1 & 2.8510 & 1.301 & 2.191 & 0.029 & [0.298, 5.404] \\
q9\_2 & -0.2123 & 1.252 & -0.170 & 0.865 & [-2.670, 2.245] \\
q9\_3 & 0.1327 & 1.263 & 0.105 & 0.916 & [-2.346, 2.612] \\
q9\_4 & 2.1627 & 1.241 & 1.743 & 0.082 & [-0.273, 4.598] \\
q9\_5 & -3.4642 & 1.315 & -2.634 & 0.009 & [-6.045, -0.883] \\
q9\_6 & -0.8716 & 1.339 & -0.651 & 0.515 & [-3.499, 1.756] \\
q9\_7 & -0.9703 & 1.276 & -0.761 & 0.447 & [-3.473, 1.533] \\
q9\_8 & 1.4518 & 1.369 & 1.060 & 0.289 & [-1.235, 4.139] \\
q9\_9 & -2.9355 & 1.307 & -2.245 & 0.025 & [-5.501, -0.370] \\
q9\_10 & 0.4525 & 1.370 & 0.330 & 0.741 & [-2.237, 3.142] \\
q9\_11 & -0.1418 & 1.399 & -0.101 & 0.919 & [-2.888, 2.604] \\
q9\_12 & -3.9899 & 1.402 & -2.845 & 0.005 & [-6.742, -1.238] \\
q9\_13 & -0.3460 & 1.883 & -0.184 & 0.854 & [-4.042, 3.350] \\
\bottomrule
\end{tabular}
\caption{OLS regression analysis results for operational maturity.}
\label{tab:ols_results_op_1}
\end{table*}

Tab.~\ref{tab:ols_results_op_2} details the impact of the total number of identified risks~(sum-Q1-Q13) on operational maturity. The coefficient is $-0.6454$ with a p-value of $0.034$, indicating a significant negative relationship. This suggests that organizations identifying a higher number of risks may experience lower operational maturity, possibly due to the complexity and resource demands of managing multiple risks simultaneously.\\

\begin{table*}[]
\centering
\begin{tabular}{>{\raggedright\arraybackslash}m{2cm} >{\raggedright\arraybackslash}m{2cm} >{\raggedright\arraybackslash}m{2cm} >{\raggedright\arraybackslash}m{2cm} >{\raggedright\arraybackslash}m{2cm} >{\raggedright\arraybackslash}m{3cm}}
\toprule
\textbf{Variable} & \textbf{Coefficient} & \textbf{Standard Error} & \textbf{t-Statistic} & \textbf{p-Value} & \textbf{95\% Confidence Interval} \\
\midrule
Constant & 39.4185 & 1.478 & 26.678 & 0.000 & [36.519, 42.318] \\
sum\_q1\_q13 & -0.6454 & 0.305 & -2.119 & 0.034 & [-1.243, -0.048] \\
\bottomrule
\end{tabular}
\caption{OLS regression analysis results for operational maturity.}
\label{tab:ols_results_op_2}
\end{table*}
The analysis presented in Tab.~\ref{tab:ols_results_op_3} examines the effect of organizations’ expectations of future AI regulation~(Q10) on operational maturity. The results show a negative relationship with a coefficient of $-2.3030$ and a p-value of $0.000$, suggesting a statistically significant impact on operational maturity. This may reflect concerns over the readiness and adaptability of operational processes in the face of impending regulatory demands.\\

\begin{table*}[]
\centering
\begin{tabular}{>{\raggedright\arraybackslash}m{2cm} >{\raggedright\arraybackslash}m{2cm} >{\raggedright\arraybackslash}m{2cm} >{\raggedright\arraybackslash}m{2cm} >{\raggedright\arraybackslash}m{2cm} >{\raggedright\arraybackslash}m{3cm}}
\toprule
\textbf{Variable} & \textbf{Coefficient} & \textbf{Standard Error} & \textbf{t-Statistic} & \textbf{p-Value} & \textbf{95\% Confidence Interval} \\
\midrule
Constant & 41.9424 & 1.364 & 30.756 & 0.000 & [39.266, 44.618] \\
q10 & -2.3030 & 0.527 & -4.369 & 0.000 & [-3.337, -1.269] \\
\bottomrule
\end{tabular}
\caption{OLS regression analysis results for operational maturity.}
\label{tab:ols_results_op_3}
\end{table*}

\subsubsection{Region and Industry-Specific Results}

This section showcases organizational and operational RAI maturity scores for specific regions and industries.

\begin{figure*}[]
    \centering
    \includegraphics[width=\linewidth]{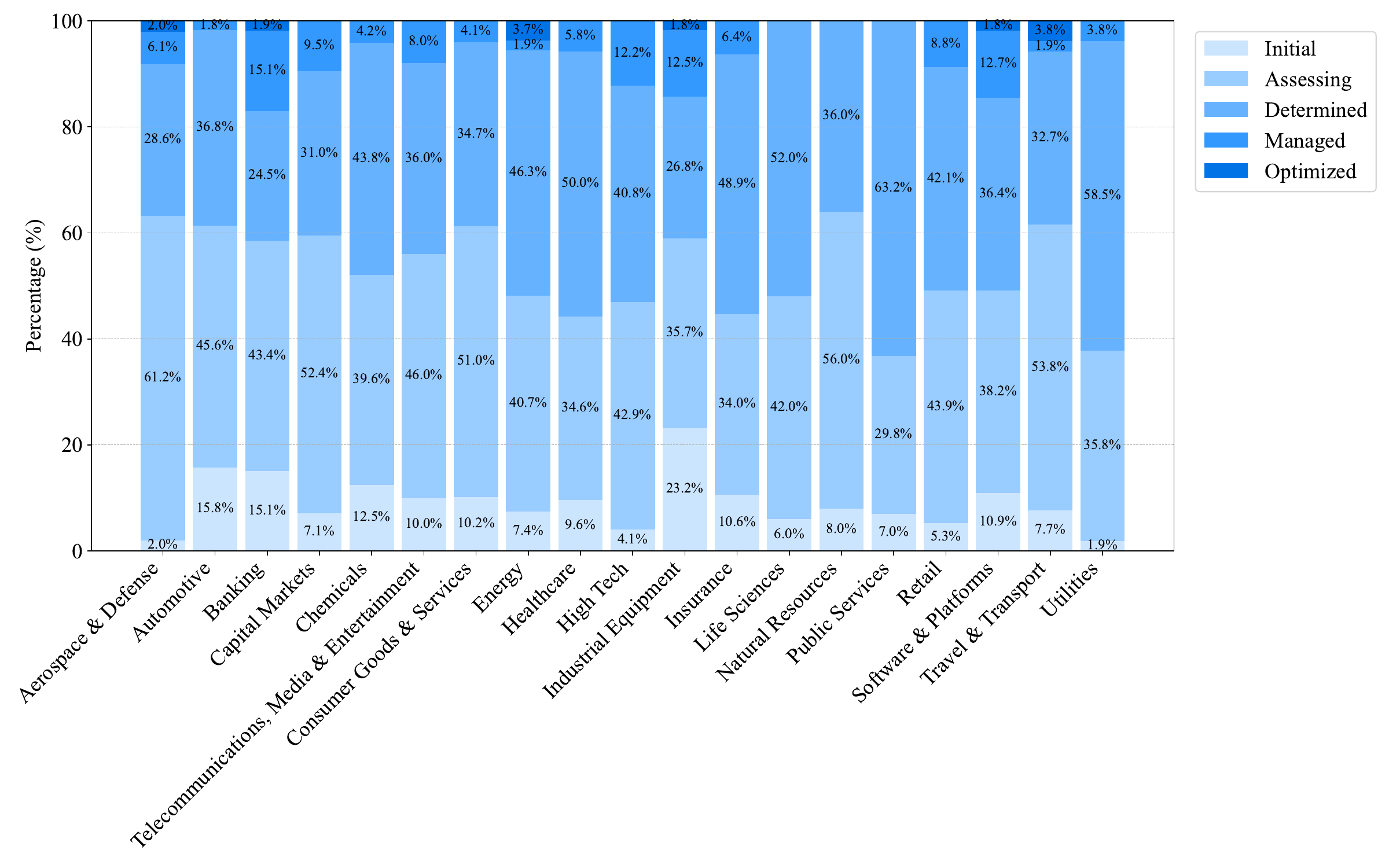}
    \caption{Operational maturity level across different industries.}
    \label{fig:op_industry}
\end{figure*}

\begin{figure*}[]
    \centering
    \includegraphics[width=\linewidth]{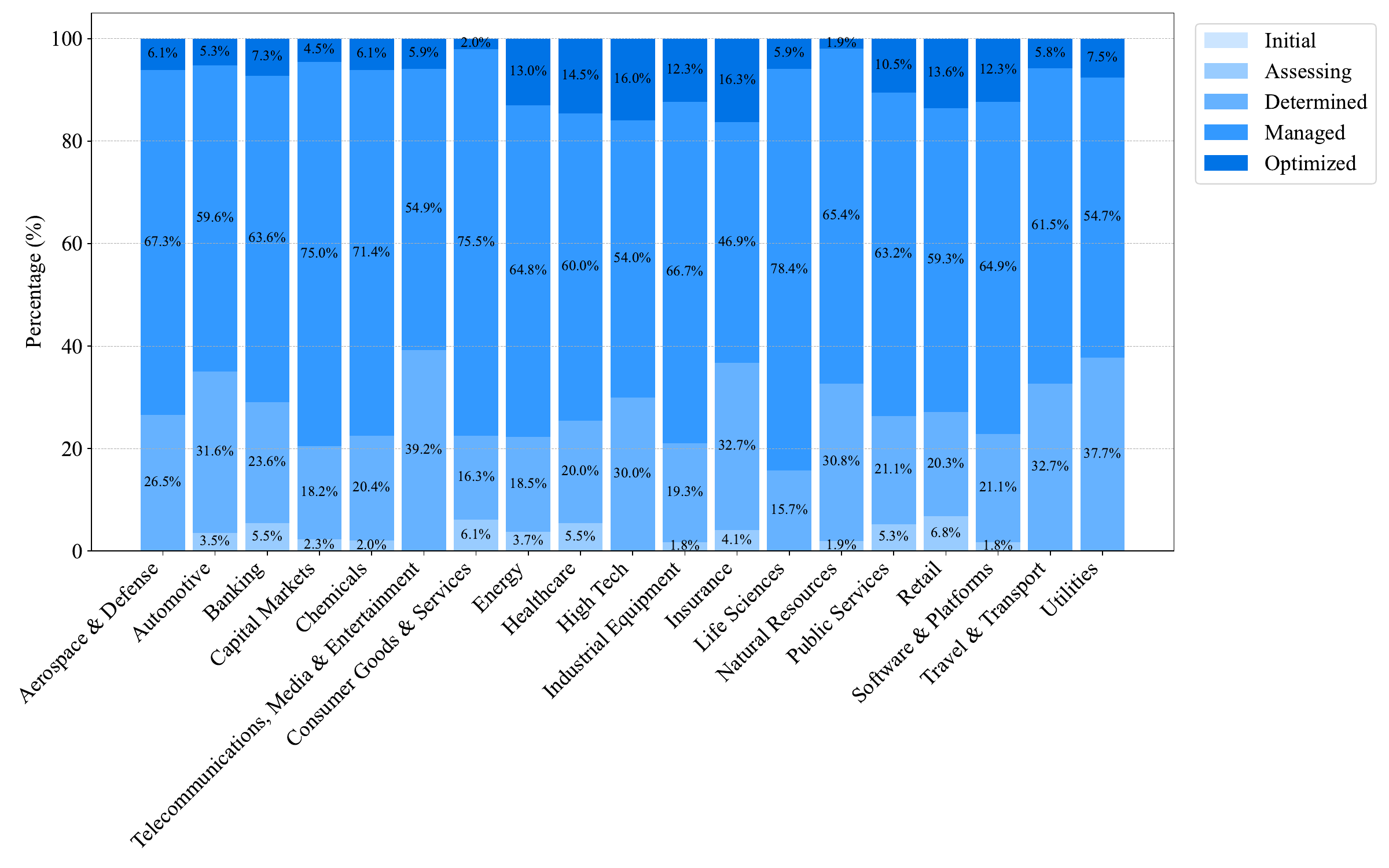}
    \caption{Organizational maturity level across different industries.}
    \label{fig:org_industry}
\end{figure*}

\begin{figure*}[]
    \centering
    \includegraphics[width=0.7\linewidth]{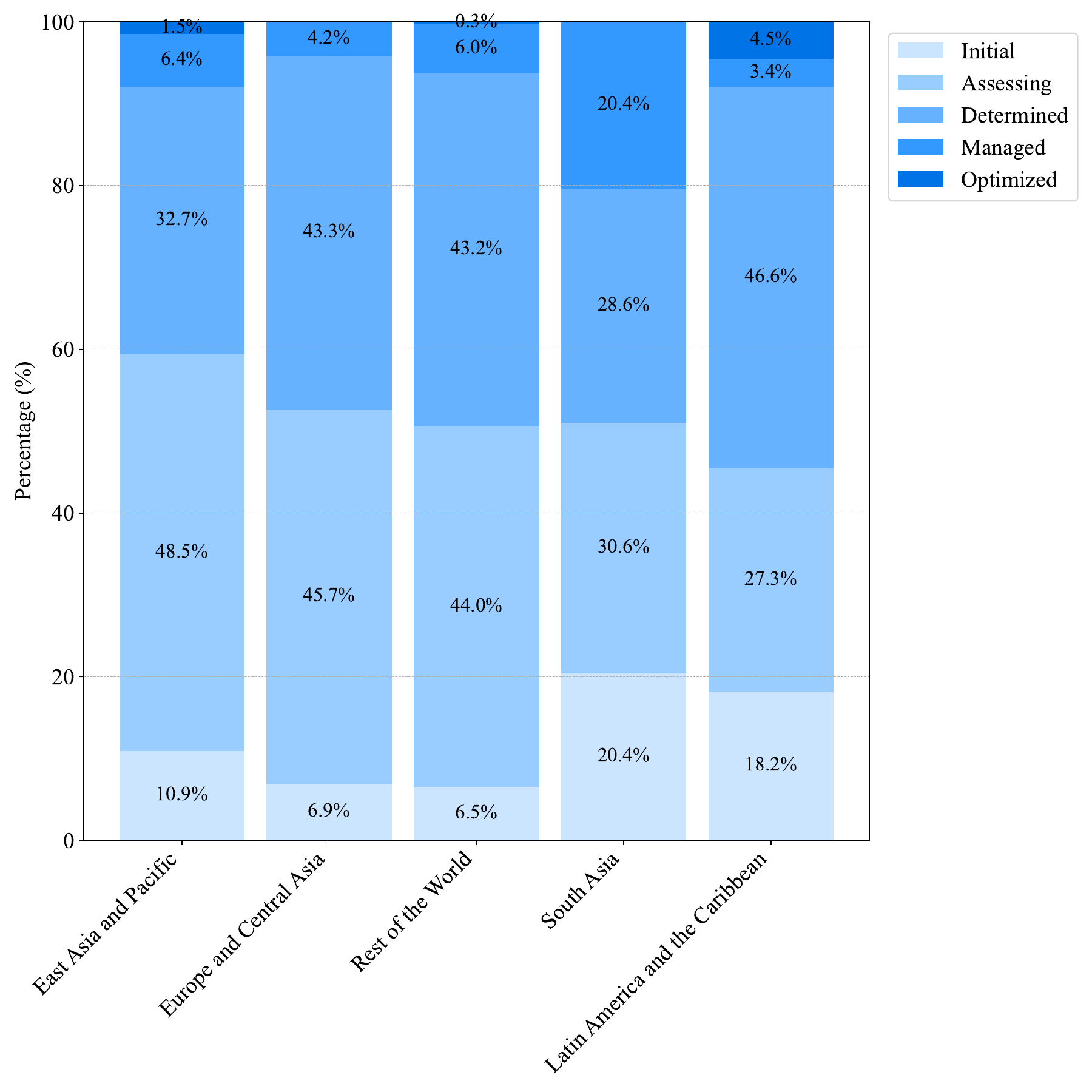}
    \caption{Operational maturity level across different regions.}
    \label{fig:org_industry}
\end{figure*}

\begin{figure*}[]
    \centering
    \includegraphics[width=0.7\linewidth]{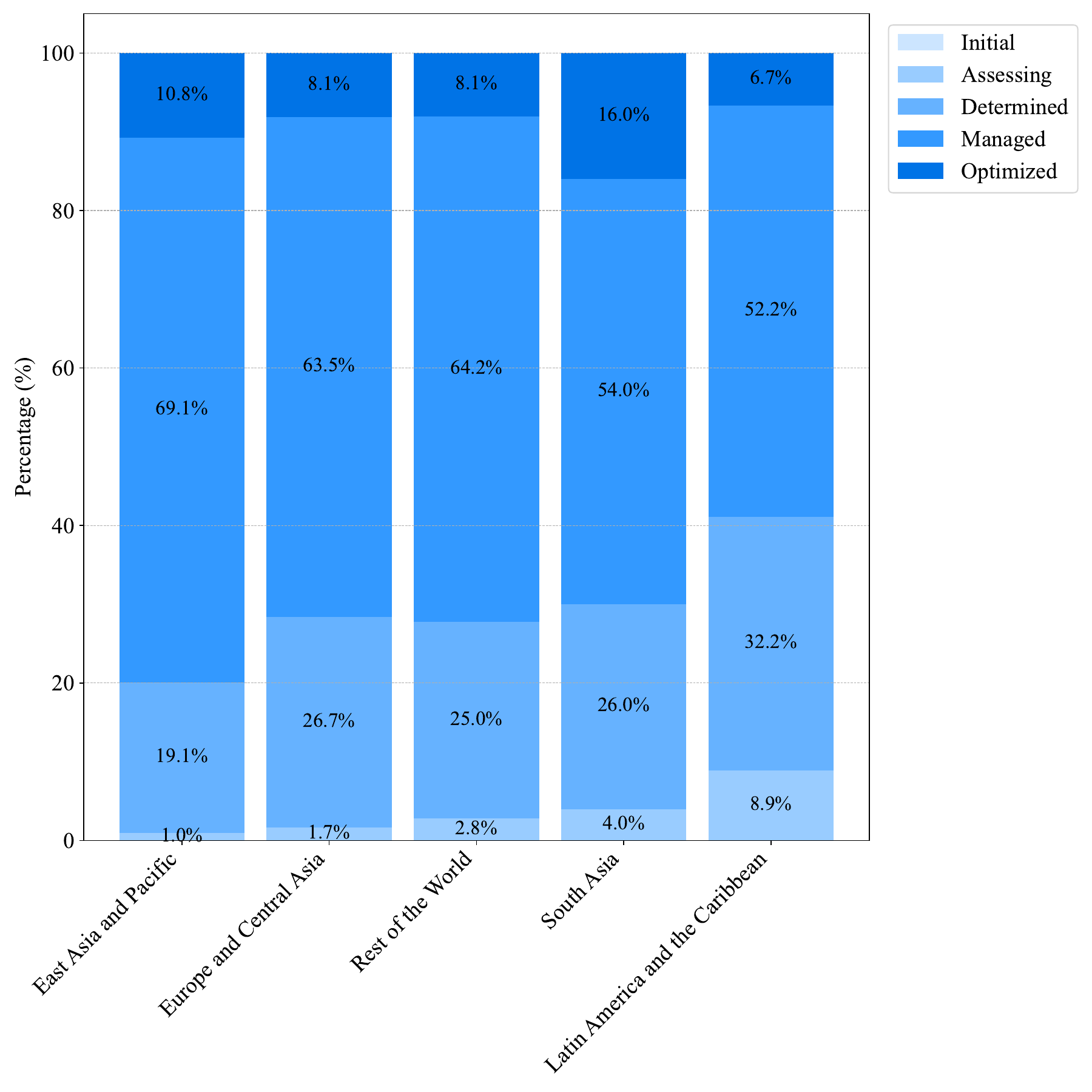}
    \caption{Organizational maturity level across different regions.}
    \label{fig:org_industry}
\end{figure*}

\subsubsection{Other Results}

\begin{comment}
\begin{figure*}[]
    \centering
    \includegraphics[width=0.7\linewidth]{imgs/genai.pdf}
    \caption{Measures to mitigate risks when using generative AI models from a third-party provider.}
    \label{fig:genai_risk}
\end{figure*}
\end{comment}

\begin{comment}
\begin{figure*}[]
    \centering
    \includegraphics[width=0.7\linewidth]{imgs/q7.pdf}
    \caption{Number of AI systems developed, used, or deployed in the surveyed organization.}
    \label{fig:q7}
\end{figure*}
\end{comment}

\begin{comment}
\begin{figure*}[]
    \centering
    \includegraphics[width=0.7\linewidth]{imgs/q8.pdf}
    \caption{Current AI adoption strategy within the organization.}
    \label{fig:q8}
\end{figure*}
\end{comment}

\begin{figure*}[]
    \centering
    \includegraphics[width=0.7\linewidth]{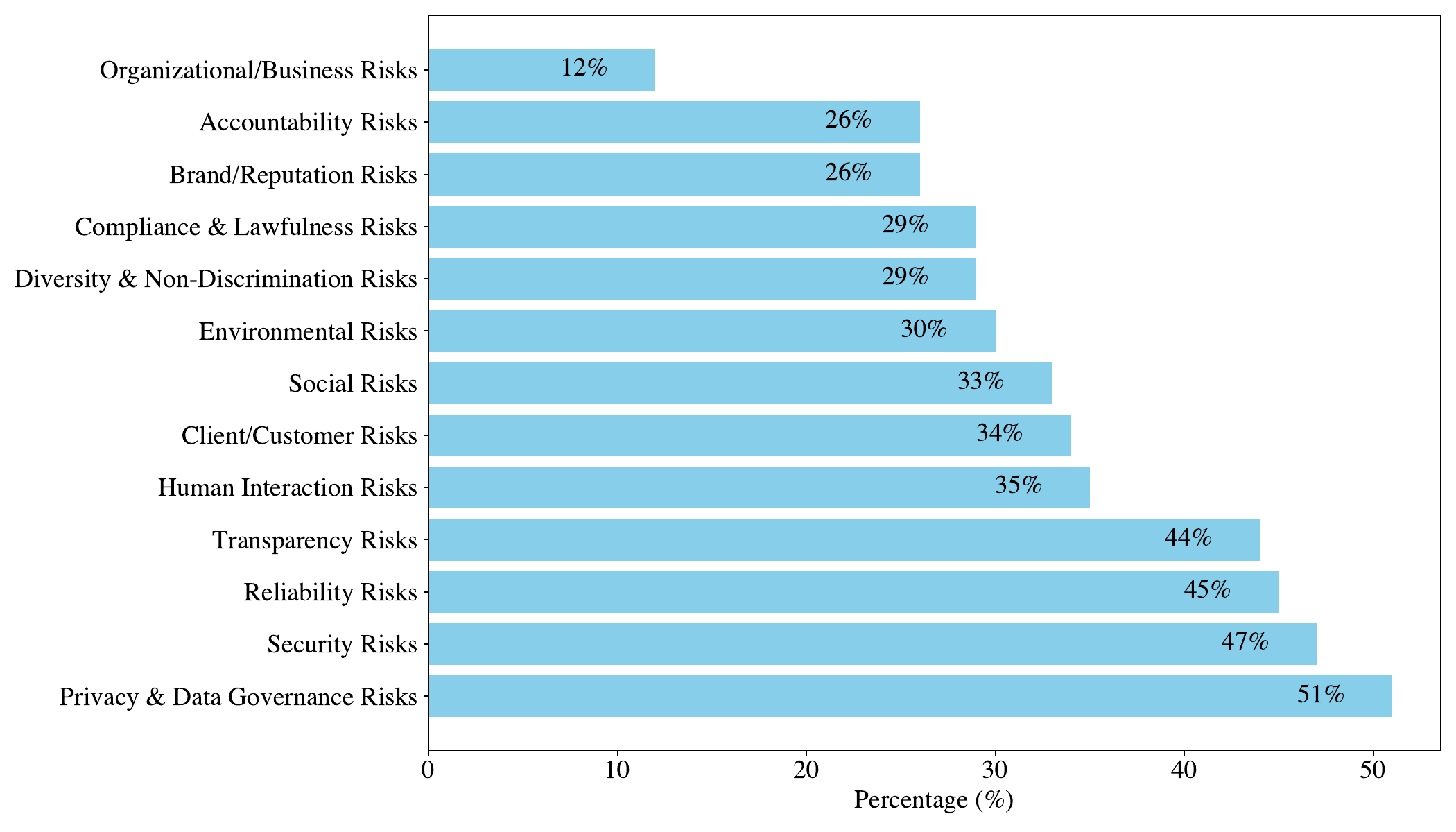}
    \caption{Risks selected by the respondents (multiple selections were possible).}
    \label{fig:q9}
\end{figure*}

\begin{figure*}[]
    \centering
    \includegraphics[width=0.7\linewidth]{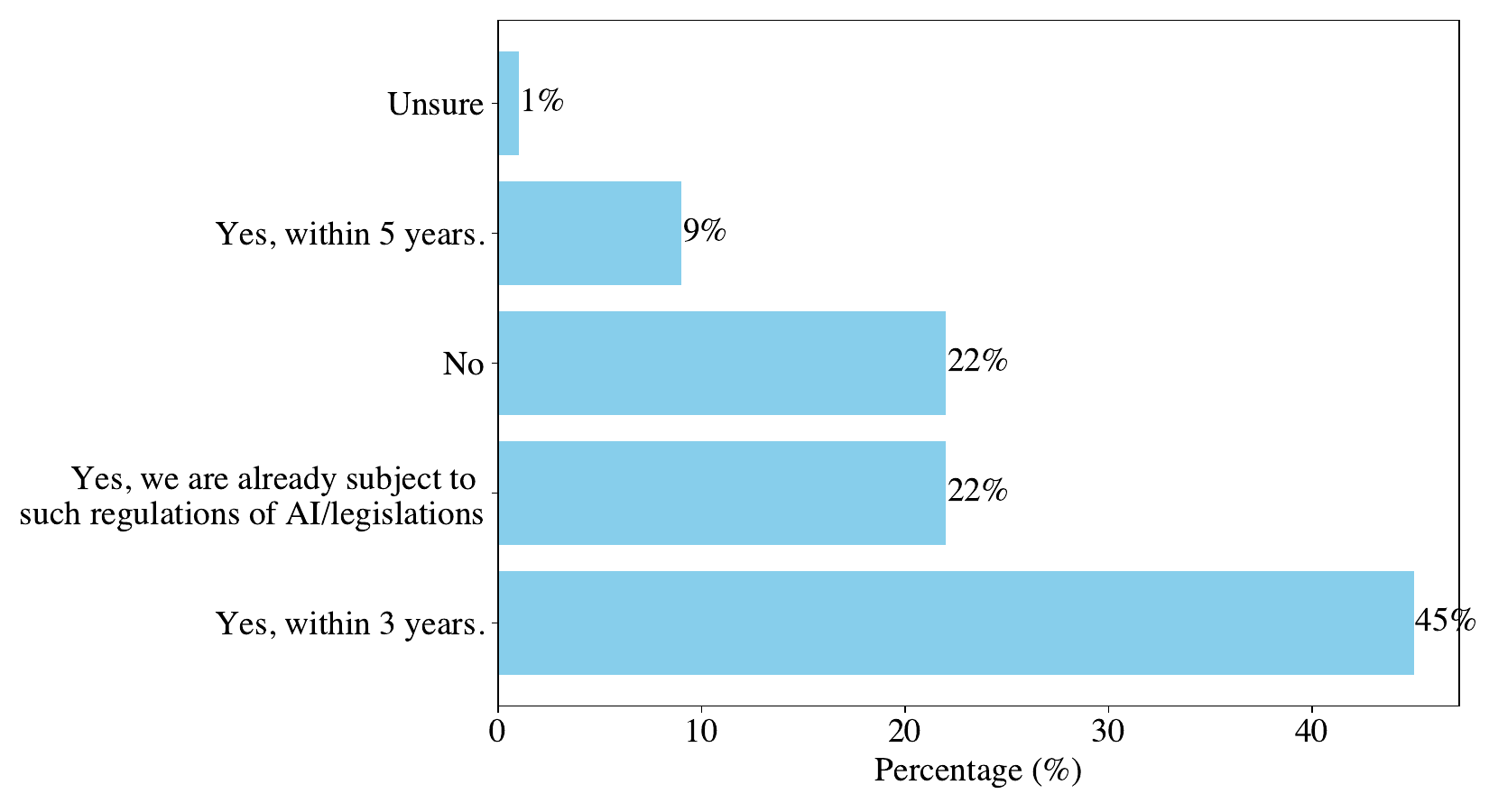}
    \caption{(Expected) regulatory exposure to specific AI regulations and legal liabilities.}
    \label{fig:q10}
\end{figure*}

\begin{comment}
\begin{figure*}[]
    \centering
    \includegraphics[width=0.7\linewidth]{imgs/q13.pdf}
    \caption{Percentage of the AI-related budget that is dedicated to RAI activities.}
    \label{fig:q13}
\end{figure*}
\end{comment}

\begin{figure*}[]
    \centering
    \includegraphics[width=0.7\linewidth]{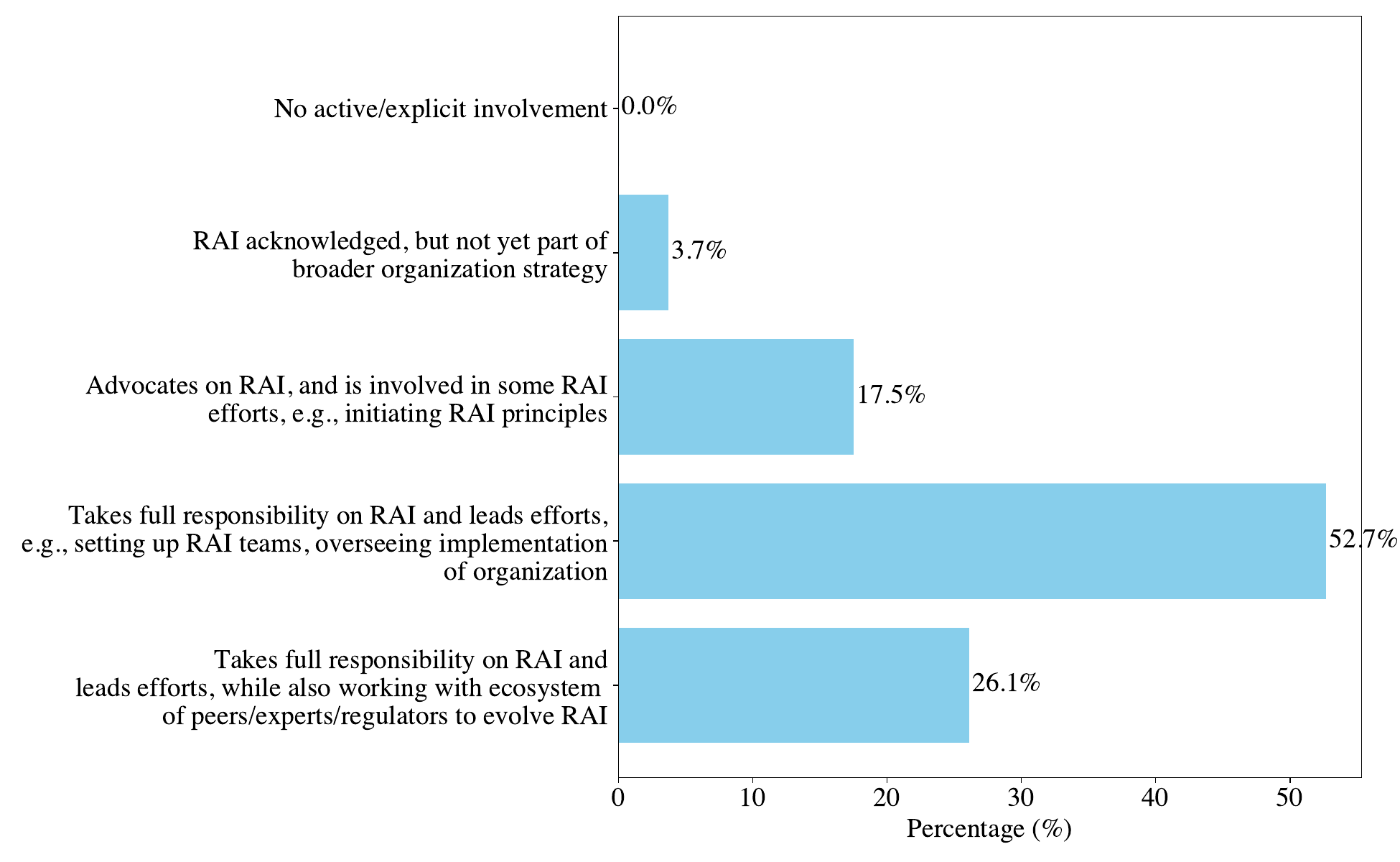}
    \caption{Involvement of CEO/Board of Directors in RAI initiatives.}
    \label{fig:q14}
\end{figure*}

\begin{figure*}[]
    \centering
    \includegraphics[width=0.7\linewidth]{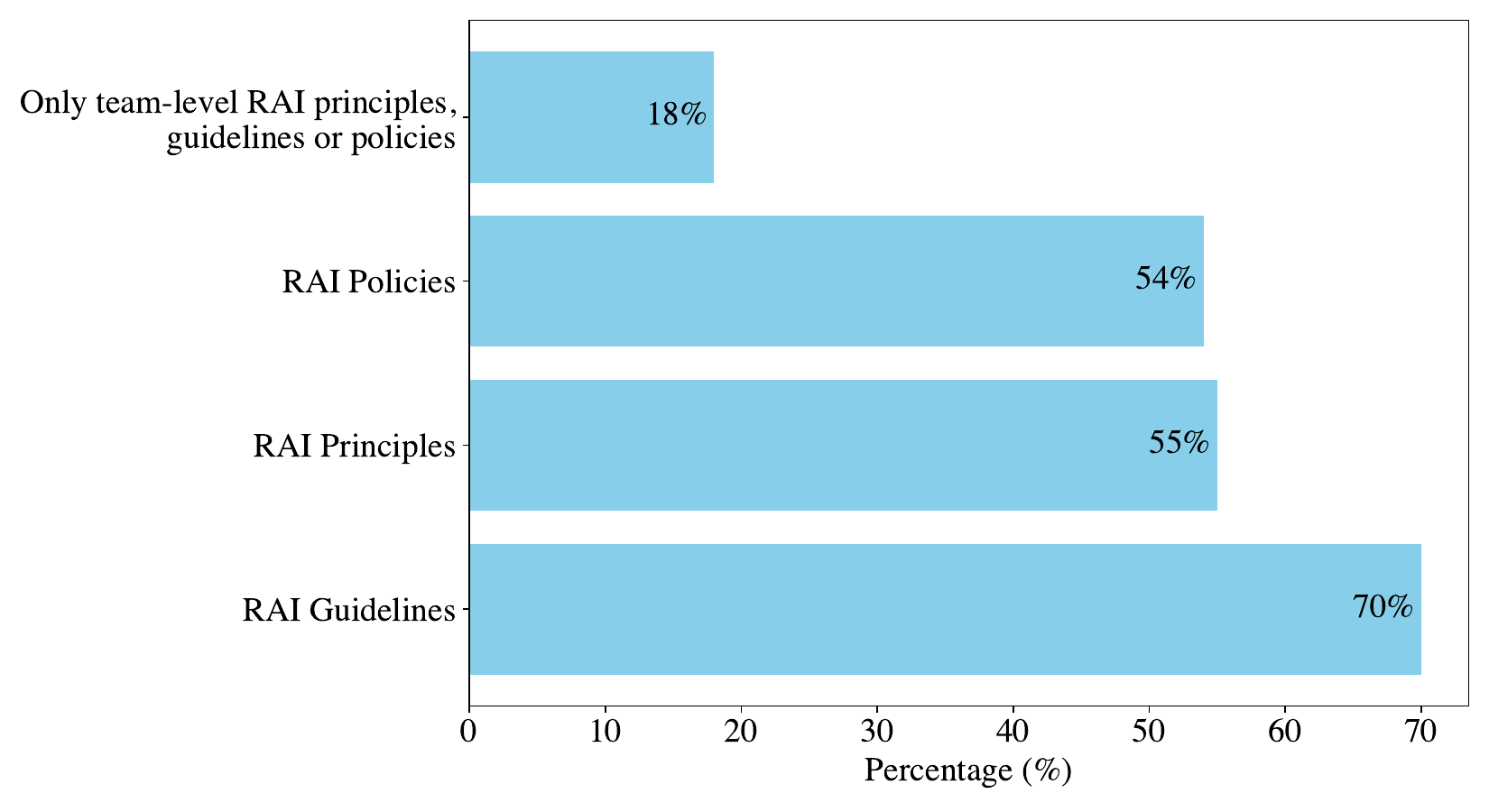}
    \caption{Indication if organization-wide RAI principles, guidelines, or policies are in place in the organization.}
    \label{fig:q16}
\end{figure*}

\begin{figure*}[]
    \centering
    \includegraphics[width=0.7\linewidth]{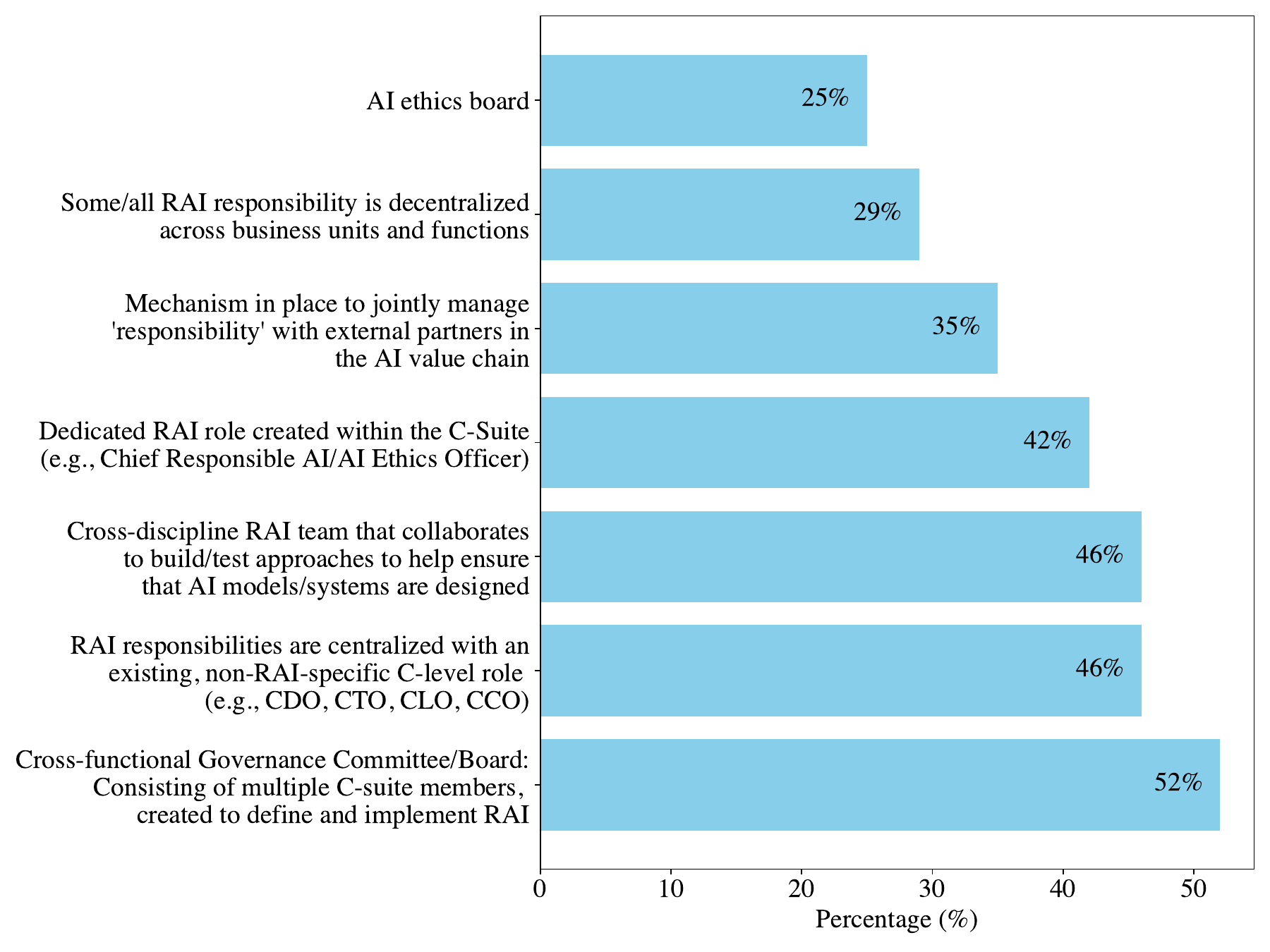}
    \caption{Dedicated RAI roles and structures that are in place in organizations.}
    \label{fig:q18}
\end{figure*}

\begin{figure*}[]
    \centering
    \includegraphics[width=0.7\linewidth]{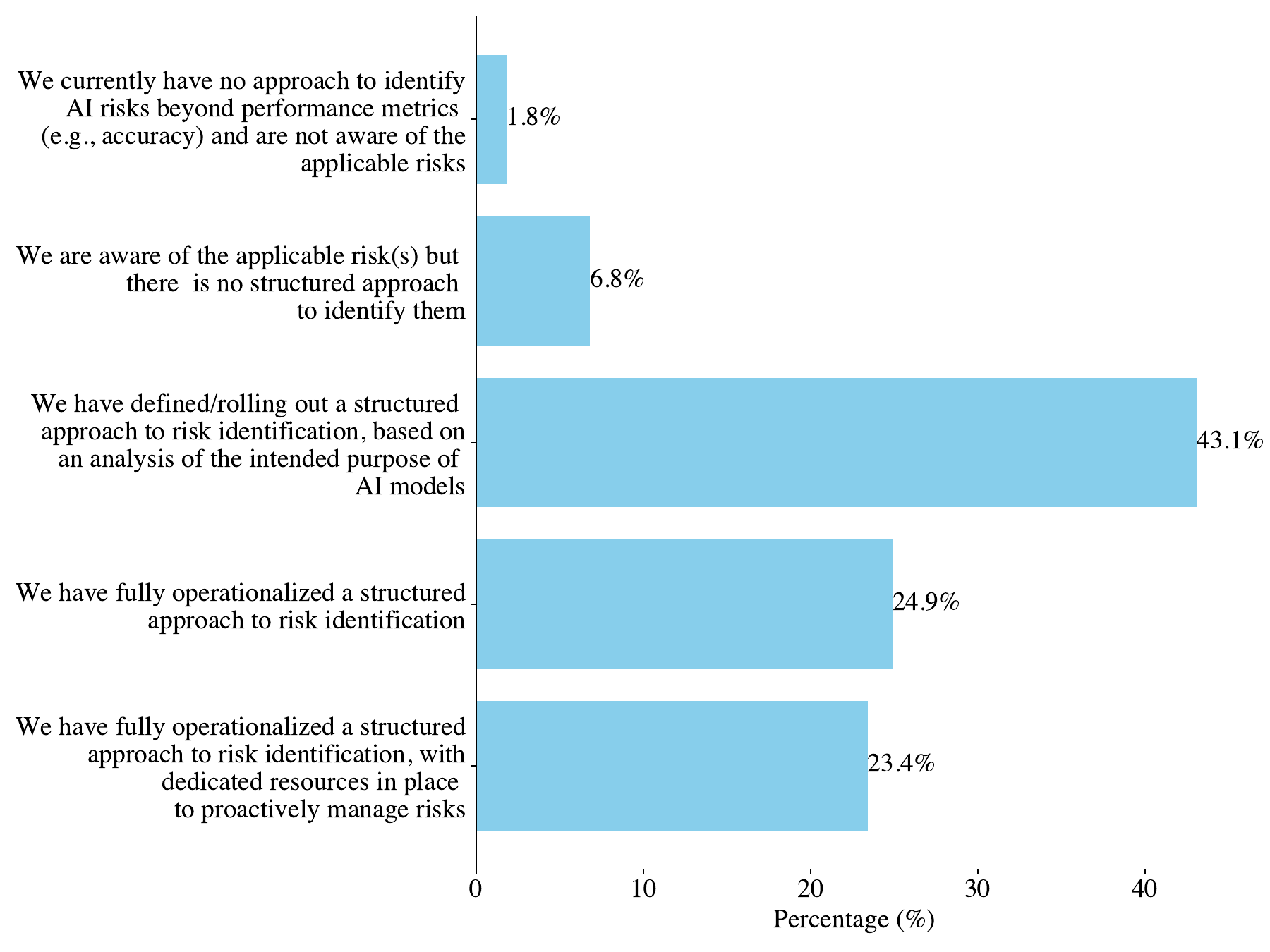}
    \caption{Risk identification processes in surveyed organizations.}
    \label{fig:q21}
\end{figure*}

\begin{figure*}[]
    \centering
    \includegraphics[width=\linewidth]{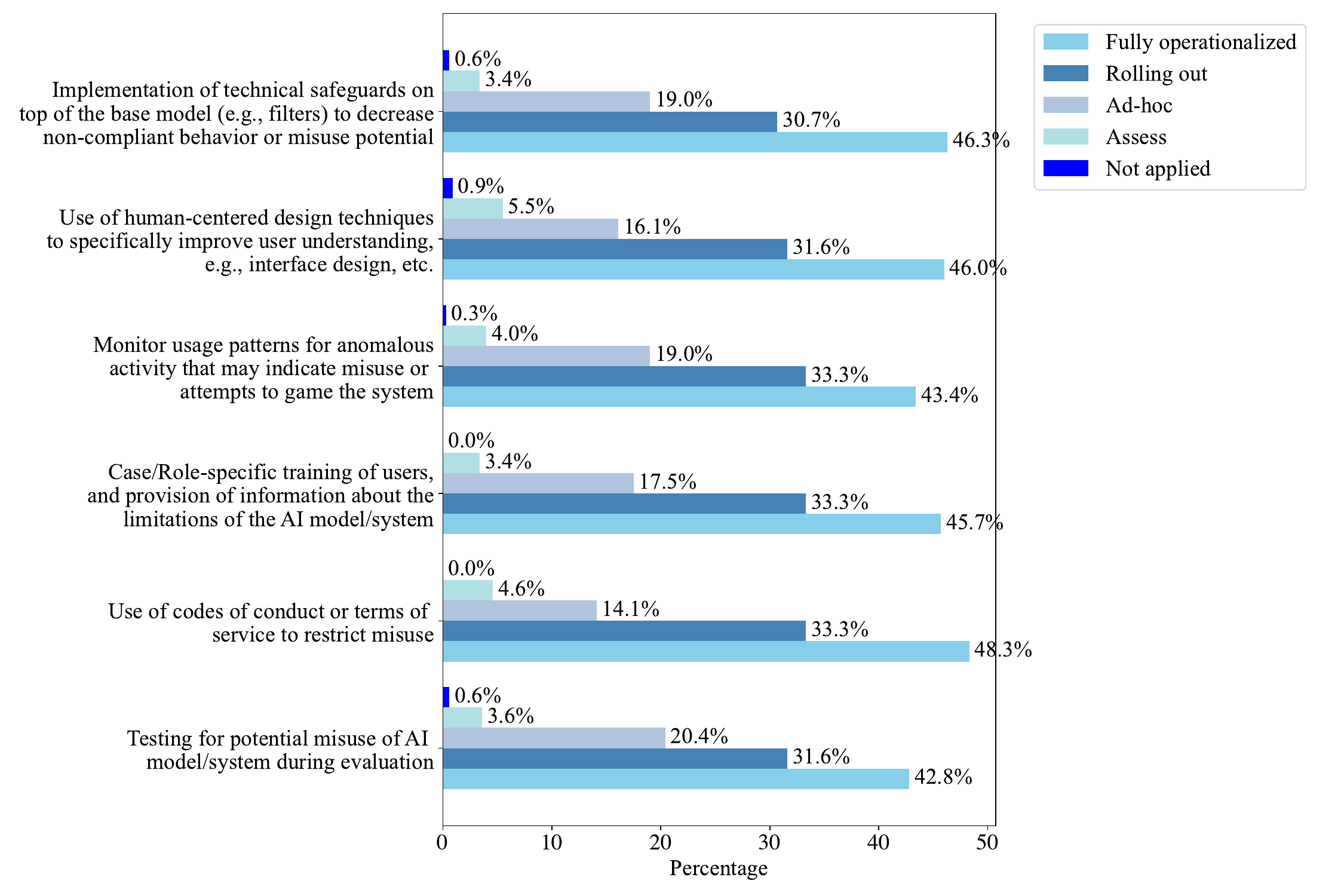}
    \caption{Implemented RAI measures to address human interaction risks.}
    \label{fig:q25}
\end{figure*}

\begin{figure*}[]
    \centering
    \includegraphics[width=\linewidth]{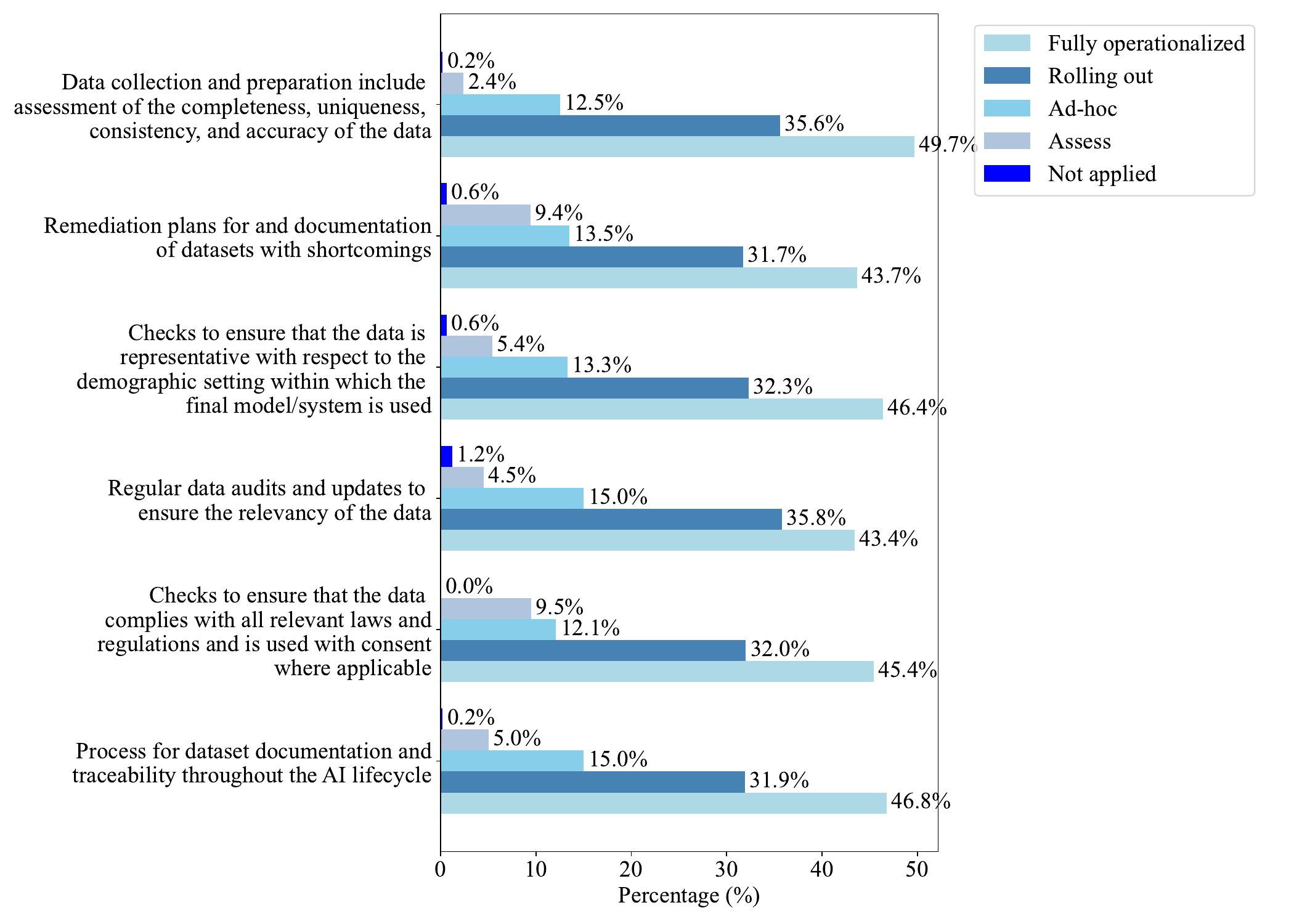}
    \caption{Implemented RAI measures to address data-related concerns.}
    \label{fig:q26}
\end{figure*}

\begin{comment}
\begin{figure*}[]
    \centering
    \includegraphics[width=\linewidth]{imgs/q24FO.pdf}
    \caption{Q24FO}
    \label{fig:q24fo}
\end{figure*}
\end{comment}

\begin{figure*}[]
    \centering
    \includegraphics[width=\linewidth]{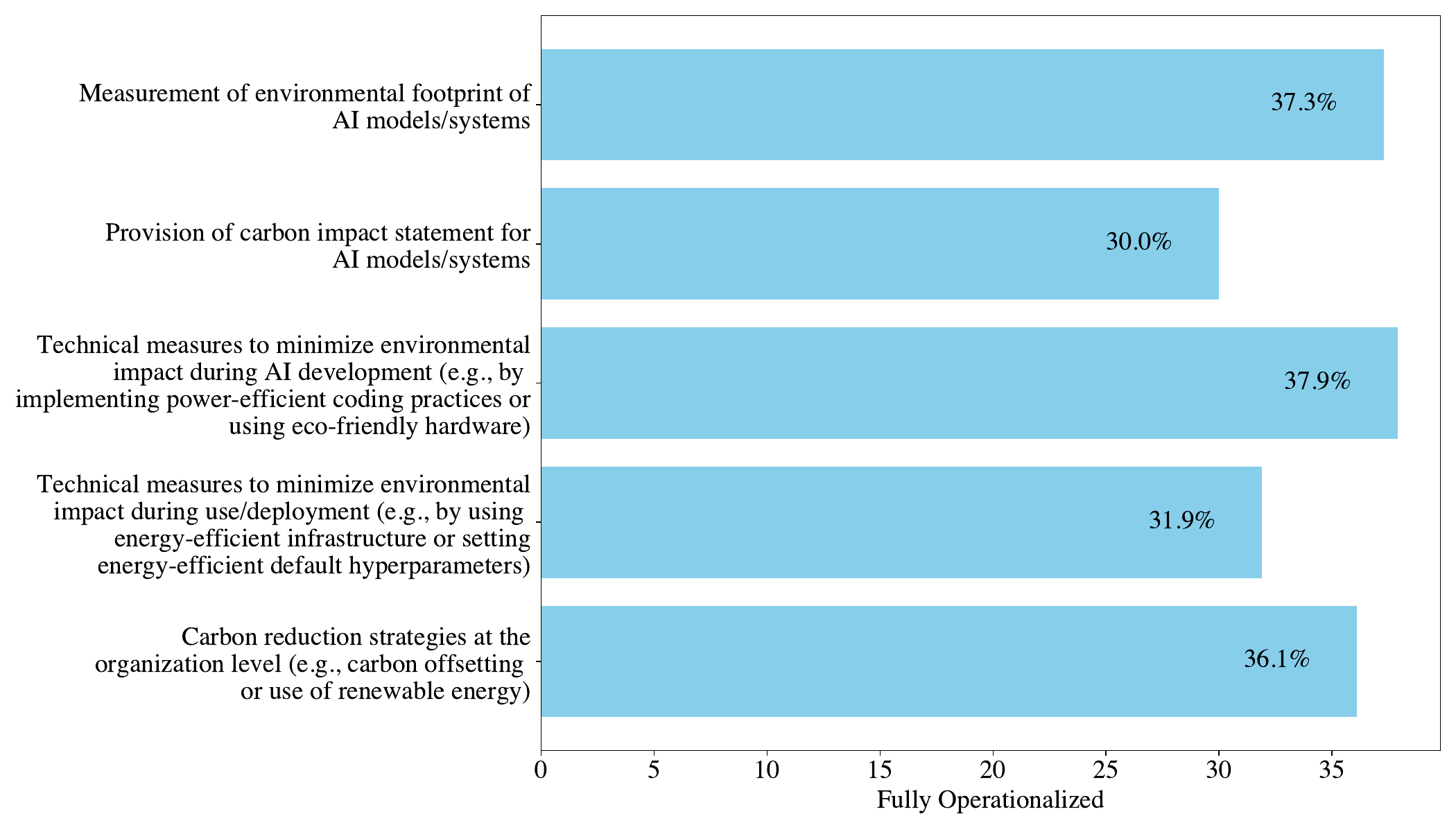}
    \caption{Implemented RAI measures to address sustainability risks.}
    \label{fig:q28fo}
\end{figure*}

\begin{figure*}[]
    \centering
    \includegraphics[width=\linewidth]{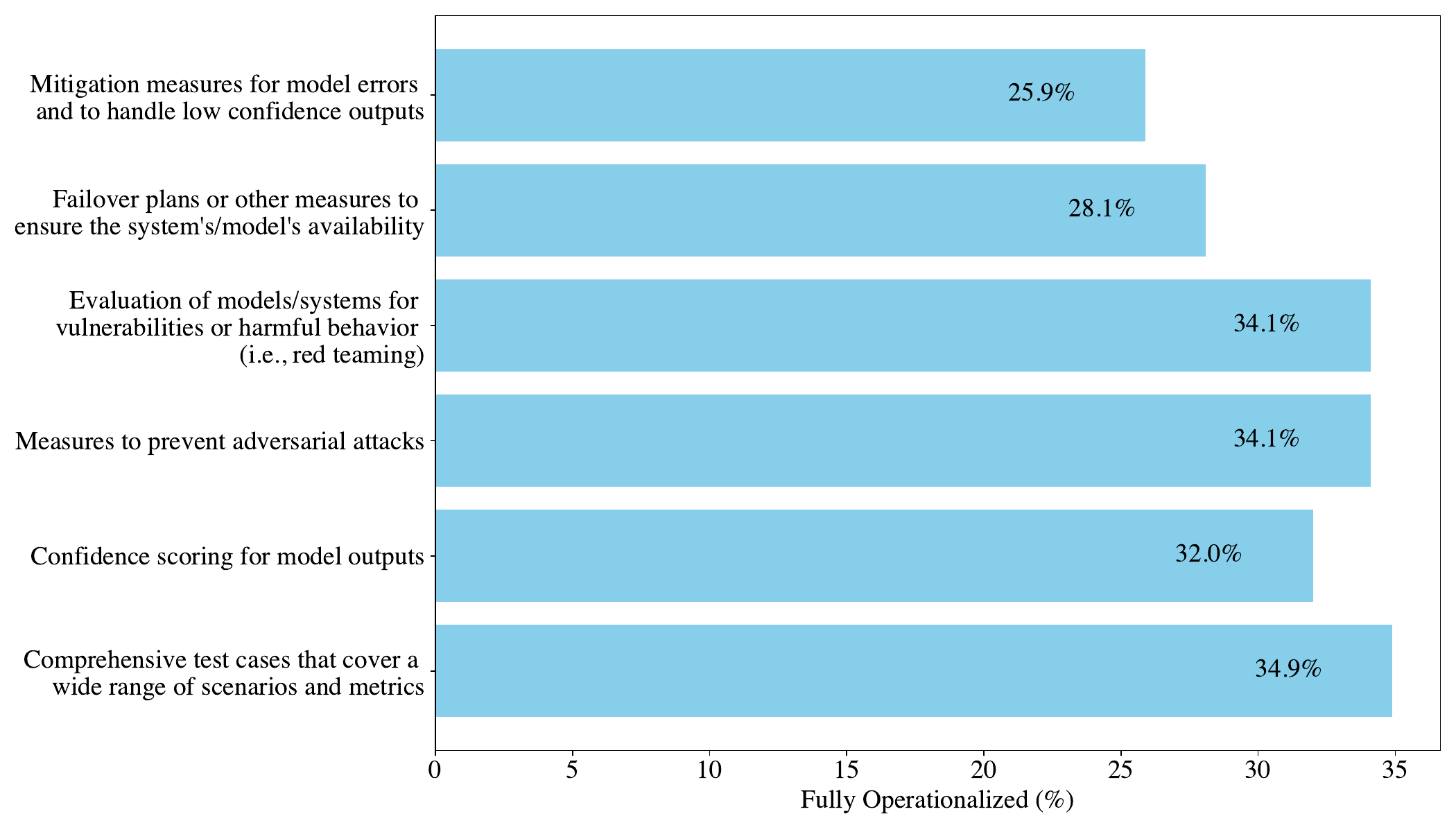}
    \caption{Implemented RAI measures to address reliability risks.}
    \label{fig:q29fo}
\end{figure*}

\begin{figure*}[]
    \centering
    \includegraphics[width=\linewidth]{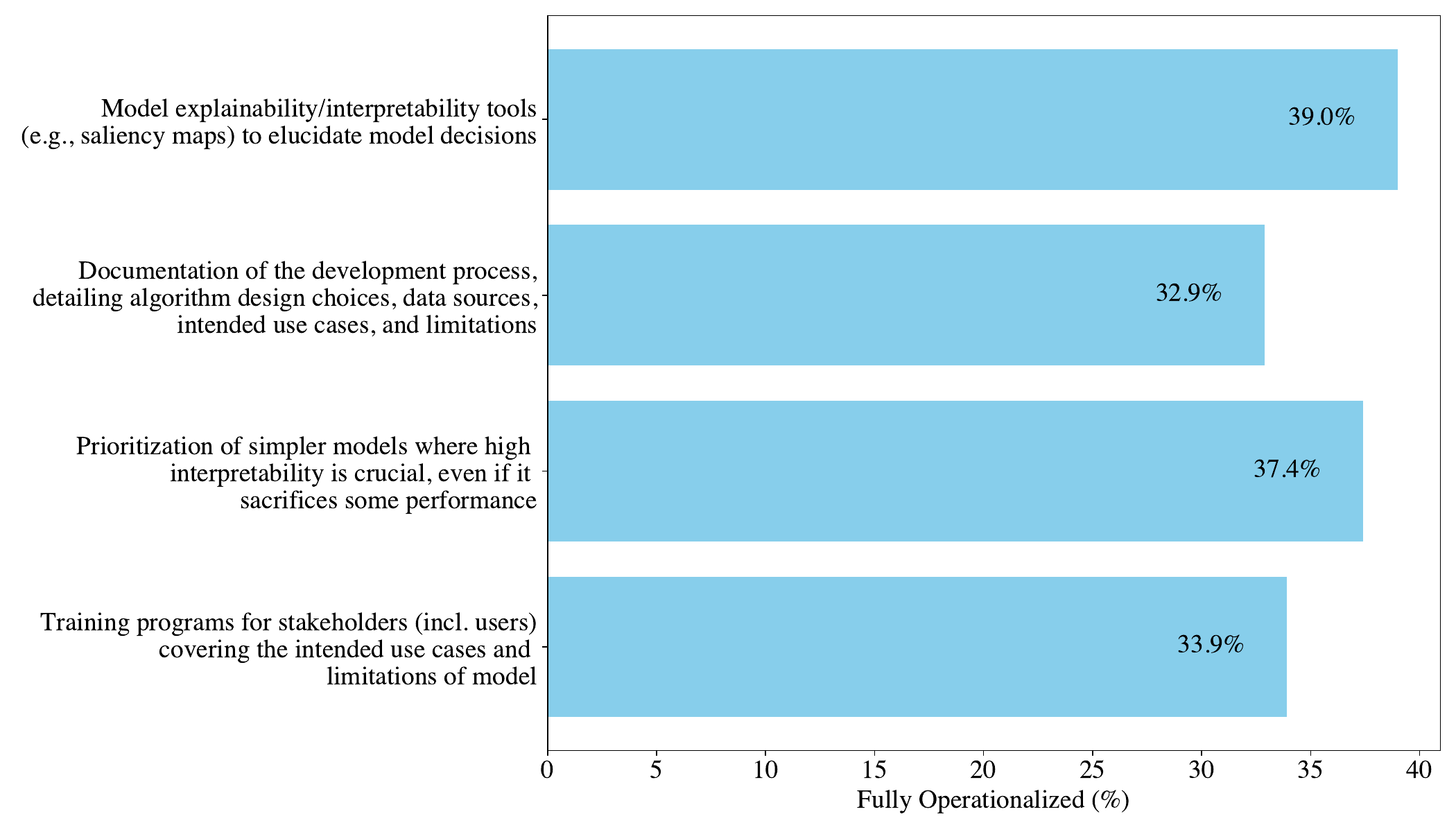}
    \caption{Implemented RAI measures to address transparency concerns.}
    \label{fig:q30fo}
\end{figure*}

\begin{figure*}[]
    \centering
    \includegraphics[width=\linewidth]{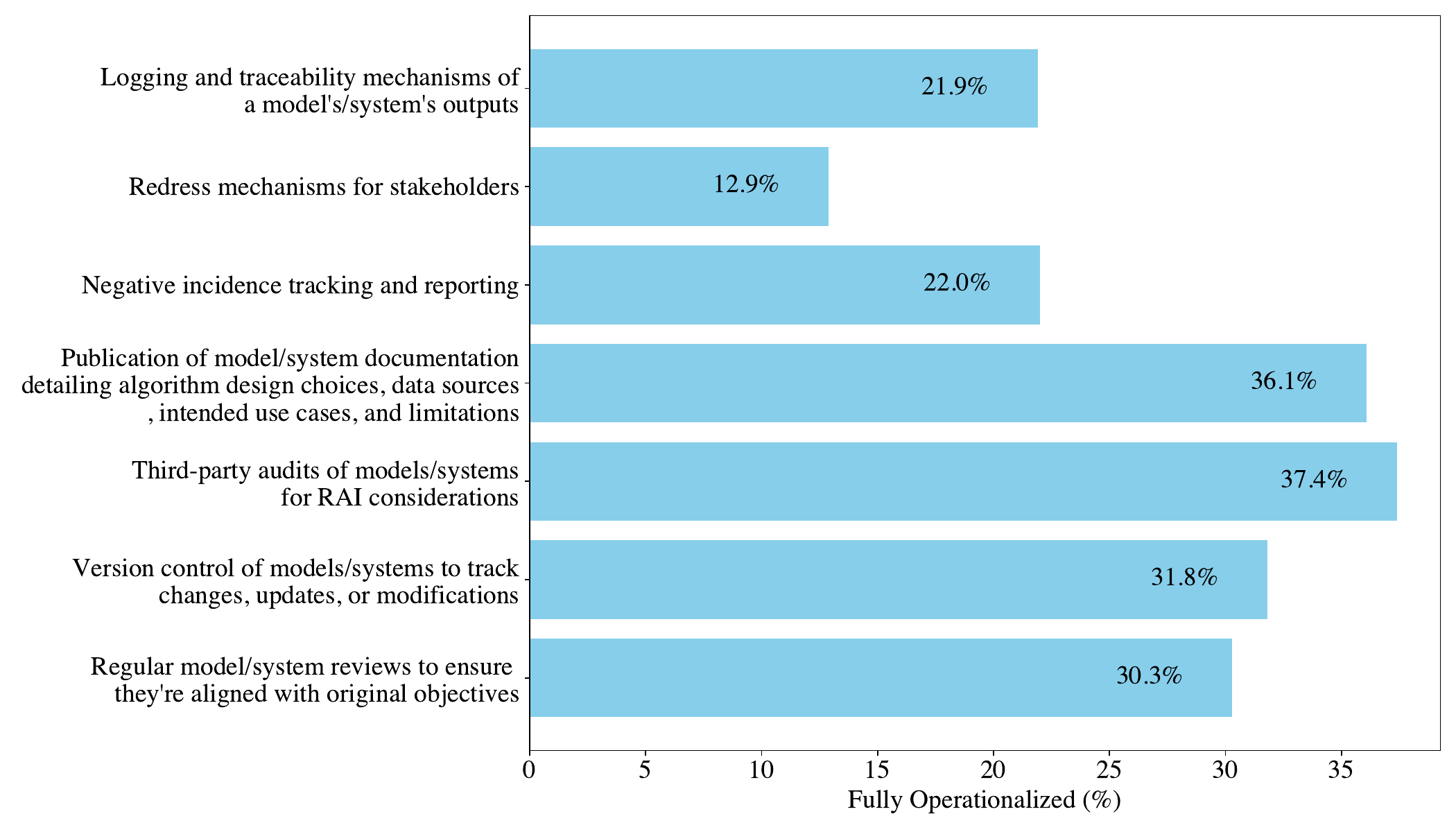}
    \caption{Implemented RAI measures to ensure accountability.}
    \label{fig:q31fo}
\end{figure*}

\begin{figure*}[]
    \centering
    \includegraphics[width=\linewidth]{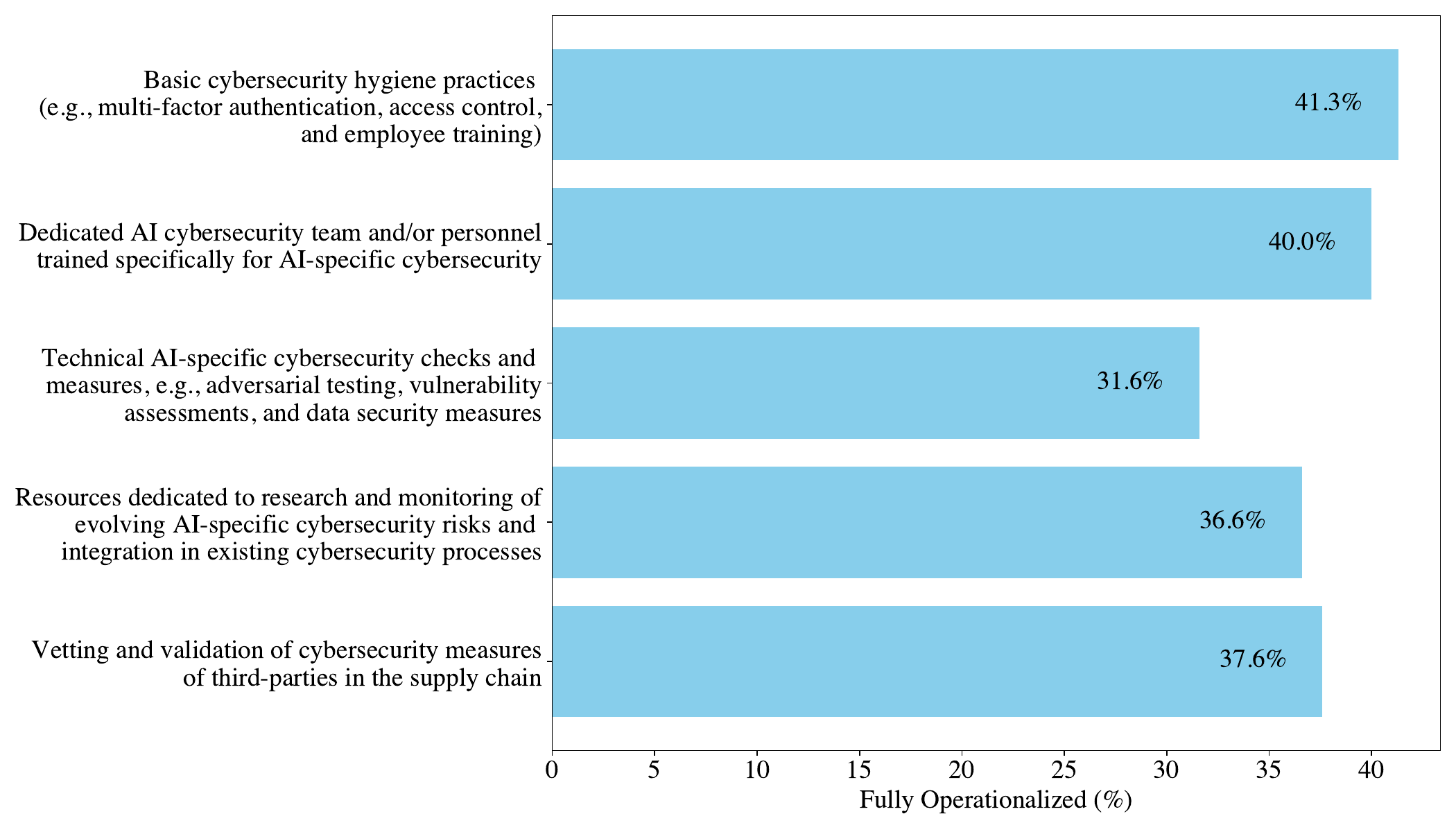}
    \caption{Implemented RAI measures to address cybersecurity risks.}
    \label{fig:q32fo}
\end{figure*}

\begin{figure*}[]
    \centering
    \includegraphics[width=\linewidth]{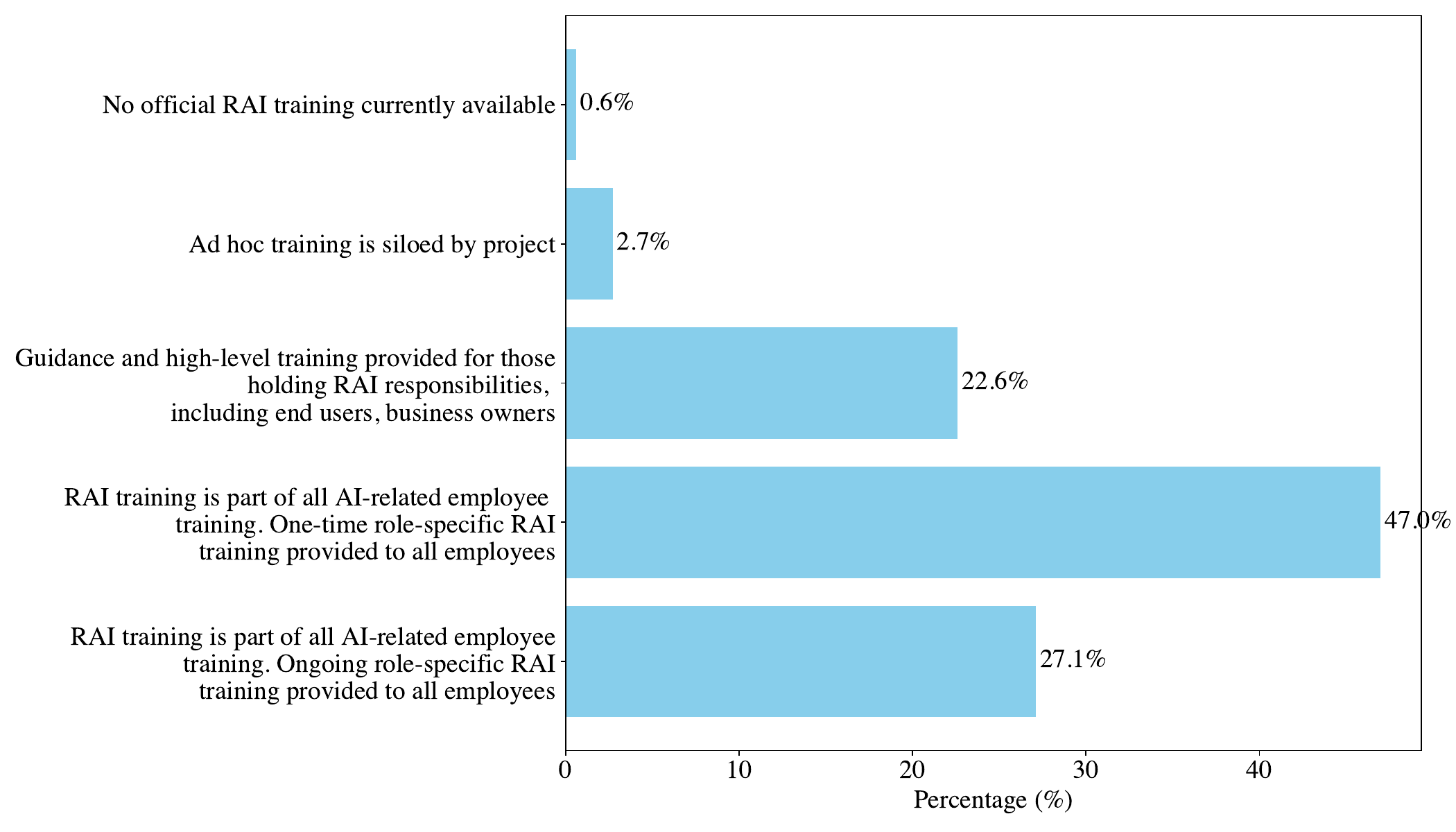}
    \caption{RAI training availability in surveyed organizations.}
    \label{fig:q33fo}
\end{figure*}

\end{document}